\documentclass[a4paper]{article}

\usepackage[utf8]{inputenc}
\usepackage[top=10mm]{geometry}
\usepackage{booktabs}
\usepackage[flushleft]{threeparttable}
\usepackage{authblk}
\usepackage{bm}
\usepackage{amsmath}
\usepackage{amssymb}
\usepackage{graphicx}
\usepackage[square,sort,numbers]{natbib}
\usepackage{slashed}
\usepackage{float}
\usepackage{longtable,booktabs}
 \allowdisplaybreaks[4]

\title{Light pseudoscalar meson and heavy meson scattering lengths to $\mathcal{O}(p^4)$ in heavy meson chiral perturbation theory}

\author[1]{Bo-Lin Huang \thanks{blhuang@pku.edu.cn}}
\author[1]{Zi-Yang Lin \thanks{lzy\_15@pku.edu.cn}}
\author[1]{Shi-Lin Zhu \thanks{zhusl@pku.edu.cn}}

\affil[1]{\textit{\small School of Physics and Center of High Energy
Physics, Peking University, Beijing 100871, China}}

\date{\today}

\begin{document}
\maketitle

\begin{abstract}
We calculate the threshold $T$ matrices of the light pseudoscalar
meson and heavy meson scattering to fourth order in heavy meson
chiral perturbation theory. We determine the low-energy constants by
fitting to the lattice QCD data points through both the perturbative
and iterated methods and obtain the physical scattering lengths in
both formalisms. The values of the scattering lengths tend to be
convergent at fourth order for most of the channels in the
perturbative method. The value of the scattering length for the
channel $DK (I=0)$, which involves the bound state
$D_{s0}^{*}(2317)$, is obtained correctly in the iterated method.
Based on the heavy diquark-antiquark symmetry, we also estimate the
meson and doubly charmed (bottom) baryon scattering lengths, and
find that the bound states can be generated with high probability in
the channels $\bar{K}\Xi_{cc}(I=0)$ and $\bar{K}\Xi_{bb}(I=0)$. We
strongly urge the LHCb Collaboration to look for the very narrow
$\bar{K}\Xi_{cc}$ state with $IJ^P=0{1\over 2}^-$ through either the
electromagnetic decay or the iso-spin violating strong decay
$\Omega_{cc} \pi$.

\begin{description}
\item[Keywords:]
Chiral perturbation theory, meson-meson interaction, scattering
lengths
\end{description}
\end{abstract}

\section{Introduction}
Since the discovery of the charm-strange meson $D_{s0}^*(2317)$ with
$J^P=0^+$ \cite{baba2003,bell2003,cleo2003}, many investigations
have been devoted to this puzzling state because its mass lies
significantly below the quark model predictions
\cite{godf1985,godf1991,dipi2001,dai1994}. The low mass puzzle of
the $D_{s0}^*(2317)$ inspired various explanations, which include
the $D^{(\ast)}K$ molecular
\cite{barn2003,chen2004,guo2006,guo2007}, the $cq\bar{s}\bar{q}$
tetraquark state \cite{chen2003,brow2004,dmit2005,brac2005}, the
conventional charmed-strange mesons with coupled-channel effects
\cite{beve2003}, and the lattice QCD simulations
\cite{bali2003,doug2003,flyn2007,liu2013,lang2014,mohl2013} (for a
detailed review see Ref.~\cite{chen2017}). In ref.~\cite{liu2013},
the low-energy interactions between light pseudoscalar mesons and
charmed pseudoscalar mesons were studied. The $S$-wave scattering
lengths of the $D\bar{K}(I=0,1)$, $D_s K$, $D\pi(I=3/2)$, and $D_s
\pi$ were calculated using L$\ddot{\text{u}}$scher's finite volume
technique in full lattice QCD. The scattering lengths of the
$D\pi(I=1/2)$, $DK(I=1,0)$, and $D_s K$ at the physical pion mass
were predicted. The attraction in the channel $DK(I=0)$ is so strong
that a pole can be generated, and the result supported the
interpretation of the $D_{s0}^{\ast}(2317)$ as a $DK$ molecule.
Since the $D_{s0}^*(2317)$ couples to the $DK$ channel strongly, the
detailed study of the S-wave $DK$ scattering will help us understand
the nature of this exotic state.

The scattering length is an important observable in the scattering
process, which encodes the information of the fundamental
interaction and can be extracted from the threshold $T$ matrix. It
is a popular method to obtain the $T$ matrix from the chiral
perturbation theory (ChPT). However, the low-energy constants in the
chiral Lagrangian need to be determined from the experimental data.
It is a good choice to take the lattice QCD data for the light
pseudoscalar meson and heavy meson interaction from refs.~
\cite{walk2009,liu2013,flyn2007,lang2014,mohl2013,bali2017,alex2019}
when the experimental data is scanty. In lattice calculations, the
light pseudoscalar meson masses are always larger than their
physical masses. Therefore, the extrapolation of the scattering
length from the nonphysical meson mass to the physical value is
necessary with the help of ChPT.

ChPT is a useful and efficient tool to study the hadronic physics
at low energies. Based on Weinberg's power-counting
\cite{wein1990,wein1991}, the chiral expressions can be organized by
the power of the small external momentum or the meson momentum (or
mass). The heavy baryon chiral perturbation theory (HBChPT) was
proposed and developed to solve the power-counting problem which
occurs in baryon ChPT \cite{gass1987,jenk1991,bern1992}. Many
achievements have been obtained in SU(2) HBChPT
\cite{ordo1992,epel1998,fett2000,kais19971,mach2011,kang2013,ente2015,kais2020}.
From refs.~
\cite{kais2001,liu20071,haid2013,huan2015,huan2017,huan20201,huan20202,huan2021},
it turned out that the calculations in SU(3) HBChPT can also lead to
reasonable predictions.

Similar to the HBChPT formalism in the meson-baryon interaction, we
use heavy meson chiral perturbation theory (HMChPT) to deal with the
charmed mesons \cite{wise1992}. In this framework, the heavy meson
$D$ is nonrelativistic. However, the $1/M$ corrections disappear
when the threshold amplitudes are considered in HMChPT. Thus, there
is no difference when the scattering lengths are either calculated
in the HMChPT or the covariant ChPT (for a review of these
approaches, see ref.~\cite{bern2008}).

In our previous paper \cite{liu2009},  we have calculated the light
pseudoscalar meson and heavy meson $S$-wave scattering lengths up to
$\mathcal{O}(p^3)$ in HMChPT and obtained a positive scattering
length for the channel $DK(I=0)$. Note that, a repulsive interaction
has a negative scattering length in our convention. Therefore, the
channel $DK(I=0)$ is attractive. However, the $T$ matrix would not
produce a pole in the perturbative calculation, which corresponds to
a bound state or resonance. Thus, the perturbative scattering length for the channel $DK(I=0)$ would not turn into a negative one. In fact, the channel $DK(I=0)$ with a bound state [$D_{s0}^{*}(2317)$] has a strong enough attractive interaction, and can lead to a negative scattering length. Then, the
calculations with various iterated methods are performed in
Refs.~\cite{guo2009,geng2010,wang2012,liu2013,alte2014,yao2015,guo2019}.
They all obtained a negative scattering length for the channel
$DK(I=0)$ based on the amplitudes below order $\mathcal{O}(p^4)$.
Nevertheless, an attractive interaction can lead to either a
positive scattering length or a negative one. Thus, the perturbative
calculation for the scattering length is also necessary. In this
paper, we will calculate the threshold $T$ matrices of the light
pseudoscalar meson and heavy meson scattering to $\mathcal{O}(p^4)$
in HMChPT in order to obtain a more precise perturbative result.
Then, we will also use the iterated method to calculate the
scattering length in order to obtain the correct scattering length
for the channel which may admit a bound state or resonance. At last,
we estimate the mesons and doubly charmed (bottom) baryon scattering
lengths based on the heavy diquark-antiquark symmetry.

This paper is organized as follows. In Sec.~\ref{lagrangian}, we
present the chiral Lagrangians up to  $\mathcal{O}(p^4)$. In
Sec.~\ref{tmatrices}, we present the Feynman diagrams and results of
the threshold $T$ matrices. In Sec.~\ref{scatteringlengths}, we
outline how to derive scattering lengths from the $T$ matrices.
Section~\ref{results} contains the presentation and discussion of
our results. The last section is a brief summary.
Appendixes~\ref{appA} and \ref{appB} contain the various scattering
lengths of the light pseudoscalar meson and doubly charmed (bottom)
baryons, respectively.

\section{Chiral Lagrangian}

\label{lagrangian} In order to calculate the light pseudoscalar
meson and heavy meson scattering lengths up to order
$\mathcal{O}(p^4)$ in heavy meson chiral perturbation theory, the
corresponding effective Lagrangian can be written as
\begin{align}
\label{eq1}
\mathcal{L}_{\text{eff}}=\mathcal{L}^{(2)}_{\phi\phi}+\mathcal{L}^{(1)}_{H
\phi}+\mathcal{L}^{(2)}_{H \phi }+\mathcal{L}^{(3)}_{H
\phi}+\mathcal{L}^{(4)}_{H\phi}.
\end{align}
The traceless Hermitian $3\times 3$ matrices $\phi$ include the
pseudoscalar Goldstone boson fields ($\pi$, $K$, $\bar{K}$, $\eta$).
The lowest-order SU(3) chiral Lagrangian for the Goldstone
meson-meson interaction take the form \cite{bora1997}
\begin{align}
\label{eq2} \mathcal{L}^{(2)}_{\phi\phi}=f^2\text{tr}(u_\mu u^\mu
+\frac{\chi_{+}}{4}),
\end{align}
where $f$ is the pseudoscalar decay constant in the chiral limit.
The axial vector quantity
$u^\mu=\frac{i}{2}\{\xi^{\dagger},\partial^\mu\xi\}$ contains odd
number meson fields. The quantity
$\chi_{+}=\xi^{\dagger}\chi\xi^{\dagger}+\xi\chi\xi$ with
$\chi=\text{diag}(m_\pi^2,m_\pi^2,2m_K^2-m_\pi^2)$ introduces
explicit chiral symmetry breaking terms. We choose the SU(3) matrix
\begin{align}
\label{eq3} U=\xi^2=\text{exp}(i\phi/f),
\end{align}
which collects the pseudoscalar Goldstone boson fields.

The lowest-order chiral Lagrangian for the heavy mesons in the heavy
quark symmetry limit is
\begin{align}
\label{eq4}
 \mathcal{L}_{H \phi}^{(1)}=-\left\langle(iv\cdot\partial H)\bar{H}\right\rangle+\left\langle H v\cdot \Gamma \bar{H}\right\rangle+g\left\langle H u_\mu \gamma^\mu \gamma_5 \bar{H}\right\rangle,
\end{align}
where $v_\mu=(1,0,0,0)$ is the heavy meson velocity, the chiral
connection $\Gamma^\mu=\frac{i}{2} [\xi^{\dagger},\partial^\mu\xi]$ contains
even number meson fields and the doublet of the ground state heavy
mesons reads
\begin{align}
\label{eq5}
H=\frac{1+\slashed{v}}{2}(P_\mu^{*}\gamma^{\mu}+iP\gamma_5),\quad
\bar{H}=\gamma^{0}H^{\dag}\gamma^{0}=(P_\mu^{*\dag}\gamma^{\mu}+iP^{\dag}\gamma_5)\frac{1+\slashed{v}}{2},
\end{align}
\begin{align}
\label{eq5} P=(D^{0},D^{+},D_s^{+}),\quad
P_\mu^{*}=(D^{0*},D^{+*},D_{s}^{+*})_\mu,
\end{align}

For the calculation of the threshold $T$ matrices, the heavy meson
Lagrangians $\mathcal{L}^{(2)}_{H \phi }$, $\mathcal{L}^{(3)}_{H \phi
}$ and $\mathcal{L}^{(4)}_{H \phi }$ in the heavy quark symmetry
limit read
\begin{align}
\label{eq8}
\mathcal{L}_{H \phi }^{(2)}=&c_0\left\langle H\bar{H}\right\rangle \text{tr}(\chi_{+})+c_1\left\langle H\chi_{+}\bar{H}\right\rangle-c_2\left\langle H\bar{H}\right\rangle \text{tr}(v\cdot u\,v\cdot u)-c_3\left\langle Hv\cdot u \, v\cdot u \bar{H}\right\rangle\nonumber\\
&-c_4\left\langle H\bar{H}\right\rangle \text{tr}(u^\mu
u_\mu)-c_5\left\langle H u^\mu u_\mu \bar{H}\right\rangle,
\end{align}

\begin{align}
\label{eq8} \mathcal{L}_{H \phi }^{(3)}=\kappa \left\langle
H[\chi_{-},v\cdot u]\bar{H}\right\rangle,
\end{align}

\begin{align}
\label{eq8} \mathcal{L}_{H \phi }^{(4)}=e_1\left\langle H\bar{H}
\right\rangle \text{tr}[(v\cdot\partial v\cdot u)(v\cdot\partial
v\cdot u)]+e_2\left\langle H(v\cdot\partial v\cdot u)(v\cdot\partial
v\cdot u)\bar{H}\right\rangle.
\end{align}
Here, the terms with quark mass in $\mathcal{L}_{H \phi }^{(4)}$ are
not considered explicitly. Indeed, for the quark mass terms, some of
the dimension four low-energy constants (LECs) simply amount to quark mass renormalizations
of some of the dimension two $c_i$. Thus, the contributions from the quark mass terms can be absorbed into the dimension two LECs. We can neglect the quark-mass terms in our calculation. The predictions for the scattering lengths will not be affected. This is a very general
phenomenon of ChPT calculations in higher orders (for details, see,
e.g., Refs.~\cite{bora1997, fett2000}).

\section{Threshold $T$ matrices}

\label{tmatrices}
\begin{figure}[t]
\centering
\includegraphics[height=13cm,width=8cm]{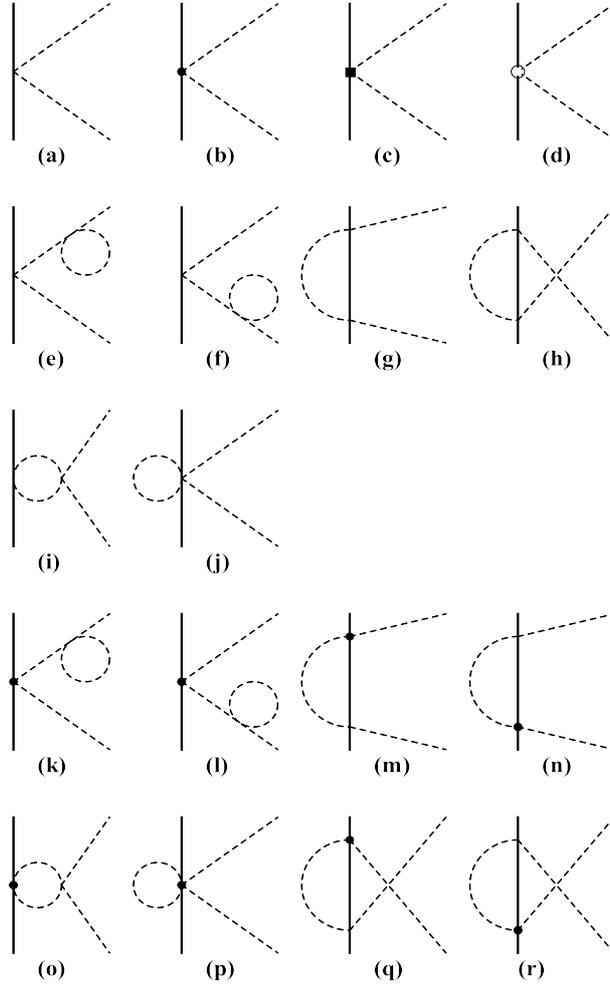}
\caption{\label{fig:feynman}Tree and non-vanishing loop diagrams in
the calculation of meson-heavy meson scattering lengths to the
fourth order in HMChPT. Dashed lines represent Goldstone bosons and
solid lines represent pseudoscalar heavy mesons. The heavy dots,
filled squares and open circles refer to vertices from
$\mathcal{L}_{ H \phi }^{(2)}$, $\mathcal{L}_{H \phi }^{(3)}$ and
$\mathcal{L}_{H \phi }^{(4)}$, respectively.}
\end{figure}

We present the explicit expressions for the threshold $T$ matrices
to $\mathcal{O}(p^4)$ in the chiral expansion. In order to obtain a
more compact representation for the $T$ matrices, the four
subfunctions are introduced,

\begin{align}
L(m_1,m_2)=\sqrt{m_1^2-m_2^2}\,\text{ln}\frac{m_1+\sqrt{m_1^2-m_2^2}}{m_2},
\end{align}
\begin{align}
L_{i}(m_1,m_2)=\sqrt{m_1^2-m_2^2}\Big(i\pi-\text{ln}\frac{m_1+\sqrt{m_1^2-m_2^2}}{m_2}\Big),
\end{align}
\begin{align}
A(m_1,m_2)=\sqrt{m_2^2-m_1^2}\,\text{arccos}\frac{m_1}{m_2},
\end{align}
\begin{align}
M(\alpha,\beta,\gamma)=\alpha m_K^4+\beta m_K^2m_\pi^2+\gamma
m_\pi^4.
\end{align}
Here, the $L_{i}(m_1,m_2)$ contains the imaginary part and comes
from the iterated diagrams (g), (m) and (n) of
Fig.~\ref{fig:feynman}. The leading-order (LO), next-to-leading
order (NLO), next-to-next-to-leading order (N2LO) and
next-to-next-to-next-to-leading order (N3LO) $T$ matrices are from
the diagrams (a), (b), (c) and (d) in Fig.~\ref{fig:feynman},
respectively. At N2LO and N3LO, we have the contributions from the
one-loop diagrams (e)-(j) and (k)-(r) of Fig.~\ref{fig:feynman},
which involve the vertices from the Lagrangians
$\mathcal{L}^{(1)}_{H\phi}$ and $\mathcal{L}^{(2)}_{H\phi}$,
respectively. The contributions from the vector charmed mesons will
not be considered explicitly in this paper. The influence of the
$D^{*}$ on the scattering lengths may be marginal \cite{yao2015}.
The description for the S-wave scattering lengths is expected to be
a good approximation in the calculation. In the following $T$ matrices of the elastic light
pseudoscalar meson and heavy meson scattering, we use $K=(K^{+},K^{0})^{T}$ and $\bar{K}=(\bar{K}^{0},K^{-})^{T}$. The
superscripts of the $T$ matrices denote the total isospin. In the
channels with an isoscalar $\eta$-meson or $D_{s}$-meson, the total
isospin is unique and does not need to be specified. We take the
renormalized (non-zero quark mass) decay constants $f_{\pi,K,\eta}$
instead of $f$ (the chiral limit value, zero quark mass). For the
construction of the one-loop and the counterterm $T$ matrics, the
scale-independent LECs $\bar{\kappa}$, $\bar{e}_1$ and $\bar{e}_2$
are used. Thus, all terms $\text{ln}(m_{\pi,K,\eta}/\mu)$ have
disappeared in our final expressions. The Gell-Mann-Okubo (GMO) relation
$m_\eta=\sqrt{(4m_K^2-m_\pi^2)/3}$ has also been used to simplify
the equations. The $c_2$ and $c_4$ ($c_3$ and $c_5$) terms are
different at N3LO only. They are strongly correlated with each
other. However, we will keep all of them because the prediction of
the physical quantity will not be affected when the correlations are
considered. Then, the $T$ matrices with
the four orders read

\begin{align}
\label{tdk1}
T_{DK}^{(1)}=&\Big\{0\Big\}_{\text{LO}}+\Big\{\frac{2(4c_0+c_2+c_4)m_K^2}{f_K^2}\Big\}_{\text{NLO}}+\Big\{\frac{m_K^2}{8\pi^2f_K^4}L_{i}(m_K,m_\pi)\Big\}_{\text{N2LO}}+\Big\{-\frac{2\bar{e}_1 m_K^4}{f_K^2}\nonumber\\
&+\frac{1}{6912\pi^2f_K^4}\{192c_0M(40,-5,7)-96c_1[M(48,5,-5)+36m_K(m_\pi^2+m_K^2)L_{i}(m_K,m_\pi)]\nonumber\\
&-6c_2M(1,-8,7)-c_3[M(1753,-8,19)+1728m_K^3 L_{i}(m_K,m_\pi)]+48c_4M(40,-5,7)\nonumber\\
&+8c_5[M(-170,7,19)-216m_K^3
L_{i}(m_K,m_\pi)]\}\Big\}_{\text{N3LO}},
\end{align}
\begin{align}
\label{tdk0}
T_{DK}^{(0)}=&\Big\{\frac{2m_K}{f_K^2}\Big\}_{\text{LO}}+\Big\{\frac{2(4c_0+4c_1+c_2+c_3+c_4+c_5)m_K^2}{f_K^2}\Big\}_{\text{NLO}}+\Big\{\frac{16\bar{\kappa} m_K^3}{f_K^2}+\frac{3m_K^2}{8\pi^2f_K^4}[m_K\nonumber\\
&-A(-m_K,m_\eta)]\Big\}_{\text{N2LO}}+\Big\{-\frac{2(\bar{e}_1+\bar{e}_2)m_K^4}{f_K^2}+\frac{1}{6912\pi^2f_K^4}\{192c_0M(40,-5,7)\nonumber\\
&+96c_1[M(204,-41,5)-36m_K(5m_K^2-3m_\pi^2)A(-m_K,m_\eta)]-6c_2M(1,-8,7)\nonumber\\
&+c_3[M(3557,8,-19)-1728m_K^3 A(-m_K,m_\eta)]+48c_4M(40,-5,7)\nonumber\\
&+8c_5[M(478,7,19)-216m_K^3 A(-m_K,m_\eta)]\}\Big\}_{\text{N3LO}},
\end{align}

\begin{align}
\label{tdkbar1}
T_{D\bar{K}}^{(1)}=&\Big\{-\frac{m_K}{f_K^2}\Big\}_{\text{LO}}+\Big\{\frac{(8c_0+4c_1+2c_2+c_3+2c_4+c_5)m_K^2}{f_K^2}\Big\}_{\text{NLO}}+\Big\{-\frac{8\bar{\kappa} m_K^3}{f_K^2}+\frac{m_K^2}{16\pi^2f_K^4}[-3m_K\nonumber\\
&-3A(m_K,m_\eta)+L(m_K,m_\pi)]\Big\}_{\text{N2LO}}+\Big\{-\frac{(2\bar{e}_1+\bar{e}_2)m_K^4}{f_K^2}+\frac{1}{6912\pi^2f_K^4}\{192c_0M(40,-5,7)\nonumber\\
&+96c_1[M(78,-23,5)+18m_K(5m_K^2-3m_\pi^2)A(m_K,m_\eta)+18m_K(m_K^2+m_\pi^2)L(m_K,m_\pi)]\nonumber\\
&-6c_2M(1,-8,7)-c_3[M(902,8,-19)+864m_K^3A(m_K,m_\eta)+864m_K^3L(m_K,m_\pi)]\nonumber\\
&+48c_4M(40,-5,7)+8c_5[M(154,7,19)+108m_K^3A(m_K,m_\eta)+108m_K^3L(m_K,m_\pi)]\}\Big\}_{\text{N3LO}},
\end{align}

\begin{align}
\label{tdkbar0}
T_{D\bar{K}}^{(0)}=&\Big\{\frac{m_K}{f_K^2}\Big\}_{\text{LO}}+\Big\{\frac{(8c_0-4c_1+2c_2-c_3+2c_4-c_5)m_K^2}{f_K^2}\Big\}_{\text{NLO}}+\Big\{\frac{8\bar{\kappa} m_K^3}{f_K^2}+\frac{3m_K^2}{16\pi f_K^4}[m_K\nonumber\\
&+A(m_K,m_\eta)+L(m_K,m_\pi)]\Big\}_{\text{N2LO}}+\Big\{\frac{(-2\bar{e}_1+\bar{e}_2)m_K^4}{f_K^2}+\frac{1}{6912\pi^2f_K^4}\{192c_0M(40,-5,7)\nonumber\\
&+96c_1[M(-174,13,5)+m_K(-90m_K^2+54m_\pi^2)A(m_K,m_\eta)+54(m_K^2+m_\pi^2)L(m_K,m_\pi)]\nonumber\\
&-6c_2M(1,-8,7)-c_3[M(4408,-8,19)+864m_K^3A(m_K,m_\eta)-2592m_K^3L(m_K,m_\pi)]\nonumber\\&+48c_4M(40,-5,7)+8c_5[M(-494,7,19)-108m_K^3A(m_K,m_\eta)+
324m_K^3L(m_K,m_\eta)]\}\Big\}_{\text{N3LO}},
\end{align}

\begin{align}
\label{tdsk}
T_{D_sK}=&\Big\{-\frac{m_K}{f_K^2}\Big\}_{\text{LO}}+\Big\{\frac{(8c_0+4c_1+2c_2+c_3+2c_4+c_5)m_K^2}{f_K^2}\Big\}_{\text{NLO}}+\Big\{-\frac{8\bar{\kappa} m_K^3}{f_K^2}-\frac{3m_K^2}{16\pi^2f_K^4}[m_K\nonumber\\
&+A(m_K,m_\eta)-L(m_K,m_\pi)]\Big\}_{\text{N2LO}}+\Big\{-\frac{(2\bar{e}_1+\bar{e}_2)m_K^4}{f_K^2}+\frac{1}{6912\pi^2f_K^4}\{192c_0M(40,-5,7)\nonumber\\
&+192c_1[M(25,9,2)-9m_K(5m_K^2-3m_\pi^2)A(m_K,m_\eta)-27m_K(m_K^2+m_\pi^2)L(m_K,m_\pi)]\nonumber\\
&-6c_2M(1,-8,7)-c_3[M(-845,-32,4)+864m_K^3A(m_K,m_\eta)+2592m_K^3L(m_K,m_\pi)]\nonumber\\
&+48c_4M(40,-5,7)+32c_5[M(64,-11,1)-27m_K^3A(m_K,m_\eta)-81m_K^3
L(m_K,m_\pi)]\}\Big\}_{\text{N3LO}},
\end{align}

\begin{align}
\label{tdskbar}
T_{D_s\bar{K}}=&\Big\{\frac{m_K}{f_K^2}\Big\}_{\text{LO}}+\Big\{\frac{(8c_0+4c_1+2c_2+c_3+2c_4+c_5)m_K^2}{f_K^2}\Big\}_{\text{NLO}}+\Big\{\frac{8\bar{\kappa} m_K^3}{f_K^2}-\frac{3m_K^2}{16\pi^2f_K^4}[m_K\nonumber\\
&-A(-m_K,m_\eta)+L_{i}(m_K,m_\pi)]\Big\}_{\text{N2LO}}+\Big\{-\frac{(2\bar{e}_1+\bar{e}_2)m_K^4}{f_K^2}+\frac{1}{6912\pi^2f_K^4}\{192c_0M(40,-5,7)\nonumber\\
&+192c_1[M(25,9,2)+9m_K(5m_K^2-3m_\pi^2)A(-m_K,m_\eta)+27m_K(m_K^2+m_\pi^2)L_{i}(m_K,m_\pi)]\nonumber\\
&-6c_2M(1,-8,7)+c_3[M(845,32,-4)+864m_K^3A(-m_K,m_\eta)+2592m_K^3L_{i}(m_K,m_\pi)]\nonumber\\
&+48c_4M(40,-5,7)+32c_5[M(64,-11,1)+27m_K^3A(-m_K,m_\eta)+81m_K^3L_{i}(m_K,m_\pi)]\}\Big\}_{\text{N3LO}},
\end{align}

\begin{align}
\label{tdpi3c2}
T_{D\pi}^{(3/2)}=&\Big\{-\frac{m_\pi}{f_\pi^2}\Big\}_{\text{LO}}+\Big\{\frac{(8c_0+4c_1+2c_2+c_3+2c_4+c_5)m_\pi^2}{f_\pi^2}\Big\}_{\text{NLO}}+\Big\{-\frac{8\bar{\kappa} m_\pi^3}{f_\pi^2}-\frac{m_\pi^2}{16\pi^2f_\pi^4}[3m_\pi\nonumber\\
&+2A(m_\pi,m_K)]\Big\}_{\text{N2LO}}+\Big\{-\frac{(2\bar{e}_1+\bar{e}_2)m_\pi^4}{f_\pi^2}+\frac{1}{768\pi^2f_\pi^4}\{128c_0M(1,1,5)+32c_1[M(1,7,16)\nonumber\\
&+12m_\pi(m_\pi^2+ m_K^2)A(m_\pi,m_K)]+4c_2(m_\pi^4-m_K^4)-c_3[m_K^4-98m_\pi^4-192m_\pi^3A(m_\pi,m_K)]\nonumber\\
&+32c_4M(1,1,5)+8c_5[M(1,1,22)+24m_\pi^3A(m_\pi,m_K)]\}\Big\}_{\text{N3LO}},
\end{align}

\begin{align}
\label{tdpi1c2}
T_{D\pi}^{(1/2)}=&\Big\{\frac{2m_\pi}{f_\pi^2}\Big\}_{\text{LO}}+\Big\{\frac{(8c_0+4c_1+2c_2+c_3+2c_4+c_5)m_\pi^2}{f_\pi^2}\Big\}_{\text{NLO}}+\Big\{\frac{16\bar{\kappa} m_\pi^3}{f_\pi^2}-\frac{m_\pi^2}{16\pi^2f_\pi^4}[6m_\pi\nonumber\\
&+A(m_\pi,m_K)-3A(-m_\pi,m_K)]\Big\}_{\text{N2LO}}+\Big\{-\frac{(2\bar{e}_1+\bar{e}_2)m_\pi^4}{f_\pi^2}+\frac{1}{768\pi^2f_\pi^4}\{128c_0M(1,1,5)\nonumber\\
&+32c_1[M(1,7,16)-6m_\pi(m_\pi^2+m_K^2)A(m_\pi,m_K)-18m_\pi(m_\pi^2+m_K^2)A(-m_\pi,m_K)]\nonumber\\
&+4c_2(m_\pi^4-m_K^4)-c_3[m_K^4-98m_\pi^4+96m_\pi^3A(m_\pi,m_K)+288m_\pi^3A(-m_\pi,m_K)]\nonumber\\
&+32c_4M(1,1,5)+8c_5[M(1,1,22)-12m_\pi^3A(m_\pi,m_K)-36m_\pi^3A(-m_\pi,m_K)]\}\Big\}_{\text{N3LO}},
\end{align}

\begin{align}
\label{tdspi}
T_{D_s\pi}=&\Big\{0\Big\}_{\text{LO}}+\Big\{\frac{2(4c_0+c_2+c_4)m_\pi^2}{f_\pi^2}\Big\}_{\text{NLO}}+\Big\{-\frac{1}{8\pi^2f_\pi^4}m_\pi^2[A(-m_\pi,m_K)+A(m_\pi,m_K)]\Big\}_{\text{N2LO}}\nonumber\\
&+\Big\{-\frac{2\bar{e}_1m_\pi^4}{f_\pi^2}+\frac{1}{384\pi^2f_\pi^4}\{64c_0M(1,1,5)+32c_1(m_\pi^2+m_K^2)[ m_K^2-6m_\pi^2\nonumber\\
&+6m_\pi A(-m_\pi,m_K)-6m_\pi A(m_\pi,m_K)]+2c_2(m_\pi^4-m_K^4)-c_3[m_K^4+96m_\pi^4\nonumber\\
&+96m_\pi^3A(m_\pi,m_K)-96m_\pi^3A(-m_\pi,m_K)]+16c_4M(1,1,5)+8c_5[M(1,1,-12)\nonumber\\
&+12m_\pi^3A(-m_\pi^2,m_K)-12m_\pi^3
A(m_\pi,m_K)]\}\Big\}_{\text{N3LO}},
\end{align}

\begin{align}
\label{tdeta}
T_{D\eta}=&\Big\{0\Big\}_{\text{LO}}+\Big\{\frac{1}{9f_\eta^2}[4(24c_0+6c_2+c_3+6c_4+c_5)m_K^2-(24c_0-12c_1+6c_2+c_3+6c_4+c_5)m_\pi^2]\Big\}_{\text{NLO}}\nonumber\\
&+\Big\{\frac{3}{16\pi f_\eta^4}m_\eta^2[L(m_\eta,m_K)+L_{i}(m_\eta,m_K)]\Big\}_{\text{N2LO}}+\Big\{\frac{1}{256\pi^2f_\eta^4}\{128c_0m_K^2(m_\eta^2+m_K^2)\nonumber\\
&+32c_1[-9m_\eta^2 m_K^2+6m_\eta^2 m_\pi^2+m_K^4+2m_\eta(5m_K^2-3m_\pi^2)L(m_\eta,m_K)+2m_\eta(-5m_K^2\nonumber\\
&+3m_\pi^2)L_{i}(m_\eta,m_K)]-4c_2m_K^4-c_3[32m_\eta^4+m_K^4+32L_{i}(m_\eta,m_K)-32L(m_\eta,m_K)]\nonumber\\
&+32c_4m_K^2(m_K^2+m_\eta^2)+8c_5[m_K^4+m_K^2m_\eta^2-4m_\eta^4-4m_\eta^3L(m_\eta,m_K)\nonumber\\
&-4m_\eta^3L_{i}(m_\eta,m_K)]\}\Big\}_{\text{N3LO}},
\end{align}

\begin{align}
\label{tdseta}
T_{D_s\eta}=&\Big\{0\Big\}_{\text{LO}}+\Big\{\frac{1}{9f_\eta^2}[8(12c_0+12c_1+3c_2+2c_3+3c_4+2c_5)m_K^2-2(12c_0+24c_1+3c_2+2c_3\nonumber\\
&+3c_4+2c_5)m_\pi^2]\Big\}_{\text{NLO}}+\Big\{\frac{3}{8\pi f_\eta^4}m_\eta^2[L(m_\eta,m_K)+L_{i}(m_\eta,m_K)]\Big\}_{\text{N2LO}}\nonumber\\
&+\Big\{\frac{1}{128\pi^2f_\eta^4}\{64c_0m_K^2(m_K^2+m_\eta^2)+32c_1[11m_\eta^2 m_K^2-6m_\eta^2 m_\pi^2+m_K^4\nonumber\\
&+2m_\eta(5m_K^2-3m_\pi^2)L_{i}(m_\eta,m_K)-2m_\eta(5m_K^2-3m_\pi^2)L_(m_\eta,m_K)]-2c_2m_K^4\nonumber\\
&-c_3[-32m_\eta^4+m_K^4-32m_\eta^3L_{i}(m_\eta,m_K)+32m_\eta^3L(m_\eta,m_K)]+16c_4m_K^2(m_\eta^2+m_K^2)\nonumber\\
&+8c_5[m_K^4+m_K^2m_\eta^2+4m_\eta^4
+4m_\eta^3L_{i}(m_\eta,m_K)-4m_\eta^3L(m_\eta,m_K)]\}\Big\}_{\text{N3LO}}.
\end{align}

\section{Scattering lengths}
\label{scatteringlengths}

The S-wave scattering length is defined through
\begin{align}
\label{scalen} a=\frac{M}{8\pi(M+m)}T_{\text{th}},
\end{align}
where $m$ and $M$ denote the light and heavy meson mass,
respectively. We take the sign convention that a repulsive
interaction has a negative scattering length. In order to obtain a
correct description for the channels which may include bound states
or resonances, the $T$ matrix must be iterated to infinite order. We
can use a Lippmann-Schwinger equation with a cutoff range scale,
denoted by $\mu$, to obtain a finite result, as done in
ref.~\cite{kais1995}. For a single channel separable potential, the
scattering length is given by
\begin{align}
\label{scalenit} a=a_{\text{Born}}(1-\frac{1}{2}\mu\,
a_{\text{Born}})^{-1}.
\end{align}
Here $a_{\text{Born}}$ includes the contributions from all the
diagrams except for the iterated diagrams (g), (m) and (n) of Fig.~\ref{fig:feynman} in this work. The contributions from the
three diagrams can be obtained through the iteration of the diagrams
(a) and (b). Thus, our calculations include all the contributions
from the tree diagrams, the renormalization diagrams and the crossed
diagrams up to forth order in the iteration of the potential. As
a result, the scattering lengths will not include explicitly the
imaginary parts.

\section{Results and discussion}
\label{results} 

Before making predictions, we have to determine the
low-energy constants. We have the values
$c_0=0.0045\,\text{GeV}^{-1}$ and $c_1=0.1112\,\text{GeV}^{-1}$ from
ref.~\cite{yao2015} through the relations $h_0=2M_0 c_0$ and
$h_1=2M_0 c_1$, where $M_0=1918$ MeV is taken for the charmed system
\cite{hofm2004}. For the other LECs, we have two fitting strategies
to determine the pertinent constants. One is using the perturbative
scattering length formula, Eq.~(\ref{scalen}), and the other is
using the iterated scattering length formula, Eq.~(\ref{scalenit}).
For the perturbative formula, the $T$ matrices are from the
Eqs.~(\ref{tdk1})-(\ref{tdseta}), and for the iterated formula, the
contributions from the iterated diagrams (g), (m) and (n) of
Fig.~\ref{fig:feynman} should be subtracted.

\subsection{Perturbative fitting}

Now we determine $c_{2,3,4,5}$, $\bar{\kappa}$, $\bar{e}_{1}$, and
$\bar{e}_2$ using the perturbative formula, Eq.~(\ref{scalen}) and
the lattice data of the five channels [$D \bar{K}(I=0)$, $D
\bar{K}(I=1)$, $D \pi(I=3/2)$, $D_s K$, $D_s \pi$] which include the
values of the scattering lengths and the masses of the pion, kaon,
$D$ and $D_s$ mesons from ref.~\cite{liu2013}, and the GMO relation
is used throughout this paper. The corresponding lattice values of
$f_{\pi}$ and $f_{K}$ are used from ref.~\cite{walk2009}, and we
always choose $f_{\eta}=1.2f_{\pi}$ in this paper. The lattice data are obtained by using the unphysical quark mass, then the mass of the light meson can  achieve a large value ($\sim 700\,\text{MeV}$). However, the chiral expressions are expanded in terms of $m_\pi/\Lambda_{\chi}$, where $m_\pi$ denotes the mass of the light meson and $\Lambda_{\chi}$ is the chiral symmetry breaking scale ($\sim 1\,\text{GeV}$). The large mass of the light meson maybe cause the problem of the convergence. Besides, the prediction for the scattering length should be independent of the inputs.   Therefore, we take three
different fits denoted as: Fit p1, Fit p2, and Fit p3 which include
the (M007, M010), (M007, M010, M020), and (M007, M010, M020, M030)
data from Refs.~\cite{liu2013,walk2009}, respectively. Thus, there
are 10, 15, and 20 data in total for Fit p1, Fit p2, and Fit p3,
respectively. The resulting LECs with the correlations between the
parameters for the three fits can be found in
Tab.~\ref{fittingresultp}. The values of the LECs from Fit p1 and Fit p2 are   roughly consistent, and the resulting LECs from Fit p3 are the large differences because the large mass of the light meson are taken. However, the results from Fit p3 are still valid because the light meson mass is still smaller than the chiral symmetry break scale ($\sim 1\,\text{GeV}$). The uncertainty for the respective
parameter is statistical, and it measures how much a particular
parameter can be changed while maintaining a good description of the
fitted data. However, the LECs cannot really vary independently of
each other because of the mutual correlations, as detailed in
refs.~\cite{doba2014,carl2016}. Thus, the large uncertainties for
LECs in our fits will not make the errors of the scattering lengths
become large, because a full error analysis requires a complete
covariance matrix which indicates the mutual correlations between
the parameters. One of the reasons for the large uncertainties is
that the number of data is small but the number of parameters is
large. We can find that the errors of Fit p3 are smaller than the
ones of Fit p1. The other reason is that the LECs correlate each
other, especially for the $c_2$ and $c_4$ ($c_3$ and $c_5$). The
absolute value of the correlation between $c_2$ and $c_4$ ($c_3$ and
$c_5$) is very close to one ($0.99$), which is consistent with the
fact that the difference of the $T$ matrix only appears at N3LO.
Therefore, we can remove one of them. However, all of them are still
retained in our calculation because the prediction of the scattering
length will not be affected when the correlations are considered.
Besides the large uncertainties and the strong correlations, the
values of the LECs are of natural size. The result
is consistent with the assumption that the contributions from the
$D^{*}$ vector mesons are marginal. We can also find that the
absolute value of $c_2$ ($c_4$) is much smaller than $c_3$ ($c_5$)
in Fit p1 and Fit p2 because the $c_2$ ($c_4$) term is suppressed by
$1/N_c$ as compared to the $c_{3}$ ($c_5$) term. In Fit p3, the
feature is inverse because the large masses of the light
pseudoscalar mesons reduce the fitting accuracy.

\begin{table*}[!t]
\centering
\begin{threeparttable}
\caption{\label{fittingresultp}Values of the various fits with the
correlations between the parameters through the perturbative
scattering lengths  formula, Eq.~(\ref{scalen}). For a detailed
description of these fits, see the main text.}
\begin{tabular}{ccccccccccccccccccc}
\midrule \toprule
 & Fit p1 & $c_2$ & $c_3$ & $c_4$ & $c_5$ & $\bar{\kappa}$ & $\bar{e}_1$ & $\bar{e}_2$ & \\
\midrule
$c_2$ ($\text{GeV}^{-1}$)&$0.53\pm 3.12$&$1.00$ &$-0.83$ & $-0.99$& $0.83$ & $0.90$ &$-0.25$&$-0.73$&\\
\midrule
$c_3$ ($\text{GeV}^{-1}$)&$-2.82\pm 3.59$&  &$1.00$&$0.90$& $-0.99$ &$-0.86$ &$0.67$ &$0.49$\\
\midrule
$c_4$ ($\text{GeV}^{-1}$)&$-0.26\pm 2.71$&  &   &$1.00$&$-0.90$&$-0.91$&$0.38$&$0.69$\\
\midrule
$c_5$ ($\text{GeV}^{-1}$)&$3.22\pm 4.44$&    &    & & $1.00$&$0.86$&$-0.67$&$-0.49$&\\
\midrule
$\bar{\kappa}$ ($\text{GeV}^{-2}$)&$0.87\pm 0.51$&    &    & & & $1.00$& $-0.33$& $-0.85$&\\
\midrule
$\bar{e}_1$ ($\text{GeV}^{-3}$)&$-0.44\pm 1.30$&    &    & & & & $1.00$& $-0.18$&\\
\midrule
$\bar{e}_2$ ($\text{GeV}^{-3}$)&$-6.15\pm 4.54$&    &     & & & & & $1.00$&\\
\midrule
$\chi^2/\text{d.o.f.}$&$\frac{3.41}{10-7}=1.14$&    &    &\\
\bottomrule
& Fit p2 & $c_2$ & $c_3$ & $c_4$ & $c_5$ & $\bar{\kappa}$ & $\bar{e}_1$ & $\bar{e}_2$ & \\
\midrule
$c_2$ ($\text{GeV}^{-1}$)&$0.96\pm 2.02$&$1.00$&$-0.76$ & $-0.99$& $0.77$ & $0.81$ &$-0.44$&$-0.53$&\\
\midrule
$c_3$ ($\text{GeV}^{-1}$)&$-3.46\pm 1.84$&  &$1.00$&$0.84$& $-0.99$ &$-0.70$ &$0.86$ &$0.21$\\
\midrule
$c_4$ ($\text{GeV}^{-1}$)&$-0.63\pm 1.68$&  &   &$1.00$&$-0.84$&$-0.82$&$0.55$&$0.48$\\
\midrule
$c_5$ ($\text{GeV}^{-1}$)&$4.08\pm 2.28$&    &    & & $1.00$&$0.71$&$-0.86$&$-0.22$&\\
\midrule
$\bar{\kappa}$ ($\text{GeV}^{-2}$)&$0.98\pm 0.29$&    &    & & & $1.00$& $-0.41$& $-0.84$&\\
\midrule
$\bar{e}_1$ ($\text{GeV}^{-3}$)&$-0.05\pm 0.66$&    &    & & & & $1.00$& $-0.11$&\\
\midrule
$\bar{e}_2$ ($\text{GeV}^{-3}$)&$-7.62\pm 2.65$&    &     & & & & & $1.00$&\\
\midrule
$\chi^2/\text{d.o.f.}$&$\frac{10.43}{15-7}=1.30$&    &    &\\
\bottomrule
& Fit p3 & $c_2$ & $c_3$ & $c_4$ & $c_5$ & $\bar{\kappa}$ & $\bar{e}_1$ & $\bar{e}_2$ & \\
\midrule
$c_2$ ($\text{GeV}^{-1}$)&$-1.05\pm 1.15$&$1.00$ &$-0.69$ & $-0.99$& $0.71$ & $0.78$ &$-0.66$&$-0.45$&\\
\midrule
$c_3$ ($\text{GeV}^{-1}$)&$-0.34\pm 0.83$&  &$1.00$&$0.76$& $-0.99$ &$-0.51$ &$0.89$ &$0.03$\\
\midrule
$c_4$ ($\text{GeV}^{-1}$)&$1.34\pm 0.93$&  &   &$1.00$&$-0.79$&$-0.76$&$0.74$&$0.40$\\
\midrule
$c_5$ ($\text{GeV}^{-1}$)&$0.20\pm 1.03$&    &    & & $1.00$&$0.54$&$-0.90$&$-0.05$&\\
\midrule
$\bar{\kappa}$ ($\text{GeV}^{-2}$)&$0.54\pm 0.17$&    &    & & & $1.00$& $-0.43$& $-0.85$&\\
\midrule
$\bar{e}_1$ ($\text{GeV}^{-3}$)&$0.99\pm 0.35$&    &    & & & & $1.00$& $-0.07$&\\
\midrule
$\bar{e}_2$ ($\text{GeV}^{-3}$)&$-5.35\pm 1.62$&    &     & & & & & $1.00$&\\
\midrule
$\chi^2/\text{d.o.f.}$&$\frac{21.76}{20-7}=1.67$&    &    &\\
\bottomrule \midrule
\end{tabular}
\end{threeparttable}
\end{table*}

\begin{figure}[!t]
\centering
\includegraphics[height=12.5cm,width=12.5cm]{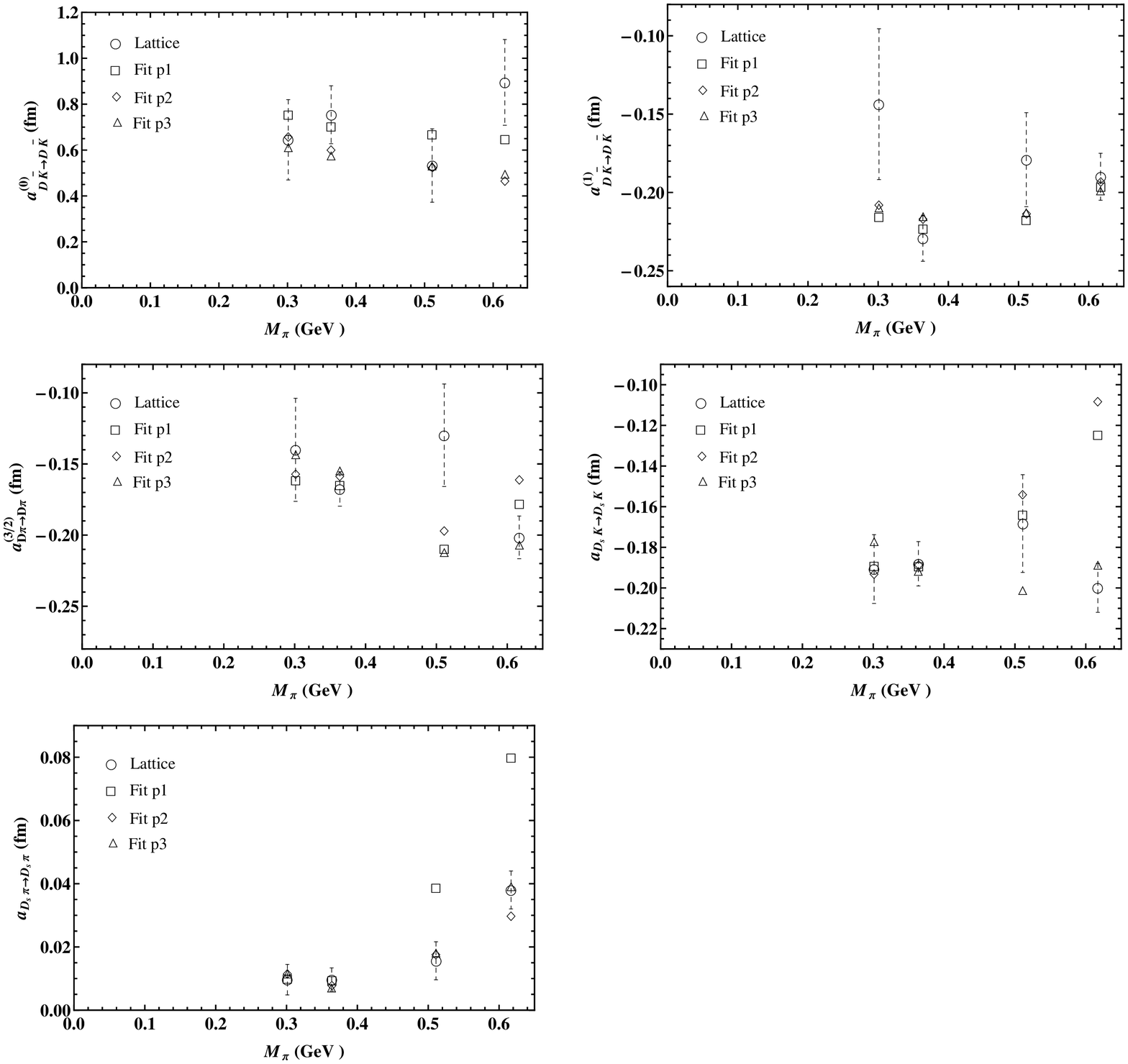}
\caption{\label{fig:fitp}Comparison of the results of the 5-channel
fits through the perturbative formula, Eq.~(\ref{scalen}), to the
lattice data of the scattering lengths from ref.~\cite{liu2013}.}
\end{figure}

The corresponding scattering lengths from the three fits are shown
in Fig.~\ref{fig:fitp}. The scattering lengths of the five channels
except for the $D \bar{K}(I=1)$ from all of the three fits are in
good agreement with the data from lattice QCD when $m_\pi\simeq
301,364\,\text{MeV}$. For the $D \bar{K}(I=1)$, the value of the
scattering length from lattice QCD has large error ($\sim 34\%$) at
$m_\pi\simeq 301\,\text{MeV}$. However, the channel at this pion
mass only has a small repulsive interaction ($\sim
-0.14\,\text{fm}$) which is far away from the nonlinear region
generated by the pole  of the iterated scattering length formula.
Thus, the large error for the data point should have a small
influence on the prediction for the strong attractive interaction.
For Fit p1, the predictions of the scattering lengths at
$m_{\pi}\simeq 511, 617\,\text{MeV}$ are away from the lattice QCD
in most data points, but they still have the same order of
magnitude. For Fit p2 and p3, the descriptions of the scattering
lengths at $m_{\pi}\simeq 511, 617\,\text{MeV}$ have been improved,
but the deviations are significantly larger than the ones at the
smaller mass of pion. The chiral expansions for the scattering
lengths are expanded in terms of the $m_{\pi,K,\eta}/\Lambda$.
Therefore, the large mass of the light pseudoscalar meson causes the
poor convergence and reduces the reliability of the predictions of
the scattering lengths. Nevertheless, the values of the
extrapolation from the low mass to high mass are consistent with the
results from lattice QCD in the same order of magnitude.

In the following, we make predictions of the scattering lengths for
the 11 channels at the physical values using the LECs determined
above. We use the values of the physical parameters:
$m_\pi=139.57\,\text{MeV}$, $m_K=493.68\,\text{MeV}$,
$f_\pi=92.07\,\text{MeV}$, $f_K=110.03\,\text{MeV}$,
$M_D=1869.66\,\text{MeV}$, $M_{Ds}=1968.35\,\text{MeV}$ from PDG
\cite{PDG2020}. The results with the three different fits are shown
in Tab.~\ref{dkscenlenp}. The errors of the scattering lengths in
the total values are estimated from the statistical errors of the
LECs using the error propagation formula with the correlations. We
can see that the errors from the LECs with large errors are not too
large to be unacceptable because of the correlations between the
parameters. The errors in the respective orders are not given
because we do not fit the scattering lengths up to the corresponding
orders, and the values of the scattering lengths from the respective
orders (i.e., the contributions exclude the lower orders) are
presented. We can find that the results tend to be convergent for most
of the channels in the three fits. The results from the three fits
are almost consistent, and the large difference does not occur. The

\begin{table}[H]
\centering
\begin{threeparttable}
\caption{\label{dkscenlenp}Predictions of the scattering lengths
through the perturbative formula, Eq.~(\ref{scalen}). The LECs are
from the three different perturbative fits, see
table~\ref{fittingresultp}. The scattering lengths are in units of
fm.}
\begin{tabular}{ccccccccccccccccccc}
\midrule \toprule
Fit p1 & $\mathcal{O}(p)$ & $\mathcal{O}(p^2)$ & $\mathcal{O}(p^3)$ & $\mathcal{O}(p^4)$ & \text{Total} & \\
\midrule
$a_{DK}^{(1)}$ &$0.00$& $0.07$ &$-0.12+0.19i$ &$0.09-0.13i$& $0.04(3)+0.07(16)i$ & \\
\midrule
$a_{DK}^{(0)}$ &$0.51$& $0.29$ & $0.77$ & $0.41$ & $1.98(93)$ & \\
\midrule
$a_{D\bar{K}}^{(1)}$ &$-0.25$& $0.18$  & $-0.49$  & $0.33$ &$-0.23(3)$&\\
\midrule
$a_{D\bar{K}}^{(0)}$ &$0.25$& $-0.04$   & $0.73$   & $-0.16$ & $0.80(6)$ &\\
\midrule
$a_{D_s K}$ &$-0.26$& $0.18$   & $-0.38$   & $0.24$ & $-0.21(7)$ & \\
\midrule
$a_{D_s \bar{K}}$ & $0.26$ & $0.18$   & $0.21+0.29i$   & $0.33+0.19i$ & $0.97(47)+0.49(22)i$ &  \\
\midrule
$a_{D\pi}^{(3/2)}$ & $-0.12$ & $0.02$  &  $-0.04$   & $0.04$ & $-0.09(3)$ &  \\
\midrule
$a_{D\pi}^{(1/2)}$ & $0.24$ & $0.02$   &  $0.02$   & $-0.01$  &  $0.27(2)$ &  \\
\midrule
$a_{D_s \pi}$ & $0.00$ &  $0.01$  &  $-0.04$  & $0.05$ & $0.02(2)$ &  \\
\midrule
$a_{D\eta}$ & $0.00$ & $0.11$   &  $0.00+0.21i$   & $0.02-0.10i$ & $0.13(8)+0.11(3)i$ & \\
\midrule
$a_{D_s \eta}$ & $0.00$ & $0.32$  &  $0.00+0.43i$   & $0.18+0.20i$ & $0.50(26)+0.62(13)i$ &  \\
\bottomrule
Fit p2 &  &   &   &   &   & \\
\midrule
$a_{DK}^{(1)}$ & $0.00$ & $0.09$ & $-0.12+0.19i$ & $0.06-0.16i$ & $0.03(3)+0.03(8)i$ &\\
\midrule
$a_{DK}^{(0)}$ & $0.51$ & $0.35$ & $0.87$   & $0.47$ &  $2.20(52)$ & \\
\midrule
$a_{D\bar{K}}^{(1)}$  & $-0.25$ & $0.22$ & $-0.55$  & $0.36$ & $-0.22(2)$ &\\
\midrule
$a_{D\bar{K}}^{(0)}$ & $0.25$ & $-0.05$   &  $0.79$  & $-0.23$ & $0.76(6)$ &\\
\midrule
$a_{D_s K}$ & $-0.26$ & $0.22$   & $-0.43$   & $0.24$ &  $-0.22(4)$ & \\
\midrule
$a_{D_s \bar{K}}$ & $0.26$ & $0.22$   &  $0.26+0.29i$  & $0.34+0.25i$ & $1.07(27)+0.54(10)i$ & \\
\midrule
$a_{D\pi}^{(3/2)}$ & $-0.12$ &  $0.03$  &  $-0.04$   & $0.03$ &  $-0.09(2)$ &\\
\midrule
$a_{D\pi}^{(1/2)}$ & $0.24$ & $0.03$   &  $0.02$   & $-0.02$ & $0.27(2)$ &\\
\midrule
$a_{D_s \pi}$ & $0.00$ & $0.01$   & $-0.04$    & $0.05$ & $0.02(2)$ & \\
\midrule
$a_{D\eta}$ & $0.00$ & $0.14$   &  $0.00+0.21i$   & $0.00-0.11i$ & $0.14(5)+0.10(1)i$ & \\
\midrule
$a_{D_s \eta}$ & $0.00$ & $0.38$   & $0.00+0.43i$  & $0.19+0.23i$ & $0.57(15)+0.65(7)i$ & \\
\bottomrule
Fit p3 &  &  &  &  &  & \\
\midrule
$a_{DK}^{(1)}$ & $0.00$ & $0.08$ & $-0.12+0.19i$ & $0.04-0.02i$ & $-0.00(2)+0.17(4)i$ & \\
\midrule
$a_{DK}^{(0)}$ & $0.51$ & $0.16$  & $0.44$ & $0.34$ & $1.44(30)$ & \\
\midrule
$a_{D\bar{K}}^{(1)}$ & $-0.25$ & $0.12$ & $-0.33$  & $0.24$ & $-0.22(2)$ &\\
\midrule
$a_{D\bar{K}}^{(0)}$ & $0.25$ & $0.04$   & $0.57$   & $-0.16$ & $0.70(4)$ &\\
\midrule
$a_{D_s K}$ & $-0.26$ &  $0.12$  & $-0.21$   & $0.20$ & $-0.15(2)$ & \\
\midrule
$a_{D_s \bar{K}}$ & $0.26$ & $0.12$   & $0.04+0.29i$   & $0.25+0.03i$ & $0.66(15)+0.33(4)i$ &  \\
\midrule
$a_{D\pi}^{(3/2)}$ & $-0.12$ & $0.02$   &  $-0.03$   & $0.06$ & $-0.07(2)$ & \\
\midrule
$a_{D\pi}^{(1/2)}$ & $0.24$ & $0.02$   &  $0.01$   &  $0.02$ & $0.28(1)$ &  \\
\midrule
$a_{D_s \pi}$ & $0.00$ & $0.01$   & $-0.04$    &  $0.05$ & $0.03(1)$ &  \\
\midrule
$a_{D\eta}$ & $0.00$ & $0.09$   &  $0.00+0.21i$   & $0.09-0.05i$ & $0.19(3)+0.16(0)i$ &  \\
\midrule
$a_{D_s \eta}$ & $0.00$ & $0.21$   & $0.00+0.43i$    & $0.16+0.11i$ & $0.37(8)+0.53(3)i$ &  \\
\bottomrule \midrule
\end{tabular}
\end{threeparttable}
\end{table}

\noindent reason is that the formulas for the predictions are linear, and the results are not very sensitive to the inputs. There exist only weak
attractions or weak repulsions in most channels. We find that the
channel $D K(I=0)$ has strong attraction but still not strong enough
to generate the well-known bound state $D_{s0}^{*}(2317)$. The value
of the scattering length should be large enough to roll over into a
negative number which is estimated to be $-1.05\,\text{fm}$ in
ref.~\cite{liu2013}. Thus, the iterated method is necessary to
describe the channel with a bound state.

\subsection{Iterated method}
Then we determine the constants using the iterated formula,
Eq.~(\ref{scalenit}). The constants include the LECs ($c_{2,3,4,5}$,
$\bar{\kappa}$, $\bar{e}_{1}$, $\bar{e}_{2}$), and also the cutoff
range scale $\mu$. The lattice data and the conventions are the same
as the perturbative method, except that the lattice QCD data of the
channel $DK(I=0)$ is also included from ref.~\cite{mohl2013}. The
data point near the physical value is used where
$m_{\pi}=156\,\text{MeV}$ and $m_K=504\,\text{MeV}$. For the values
of the $M_D$, $f_\pi$ and $f_K$, we take their physical parameters from PDG, see above. The difference is $\sim 12\%$ for $m_\pi$, and
only $\sim 2\%$ for $m_K$. Thus, the deviation from these parameters
is expected to be small. We also take three different fits indicated
as: Fit u1, Fit u2, and Fit u3 which correspond to the data of the
five channels used in perturbative method. Therefore, there are 11,
16, and 21 data in total for Fit u1, Fit u2, and Fit u3,
respectively. The resulting LECs with the correlations for the three
fits can be found in Tab.~\ref{fittingresultu}.

\begin{figure}[!t]
\centering
\includegraphics[height=12.5cm,width=12.5cm]{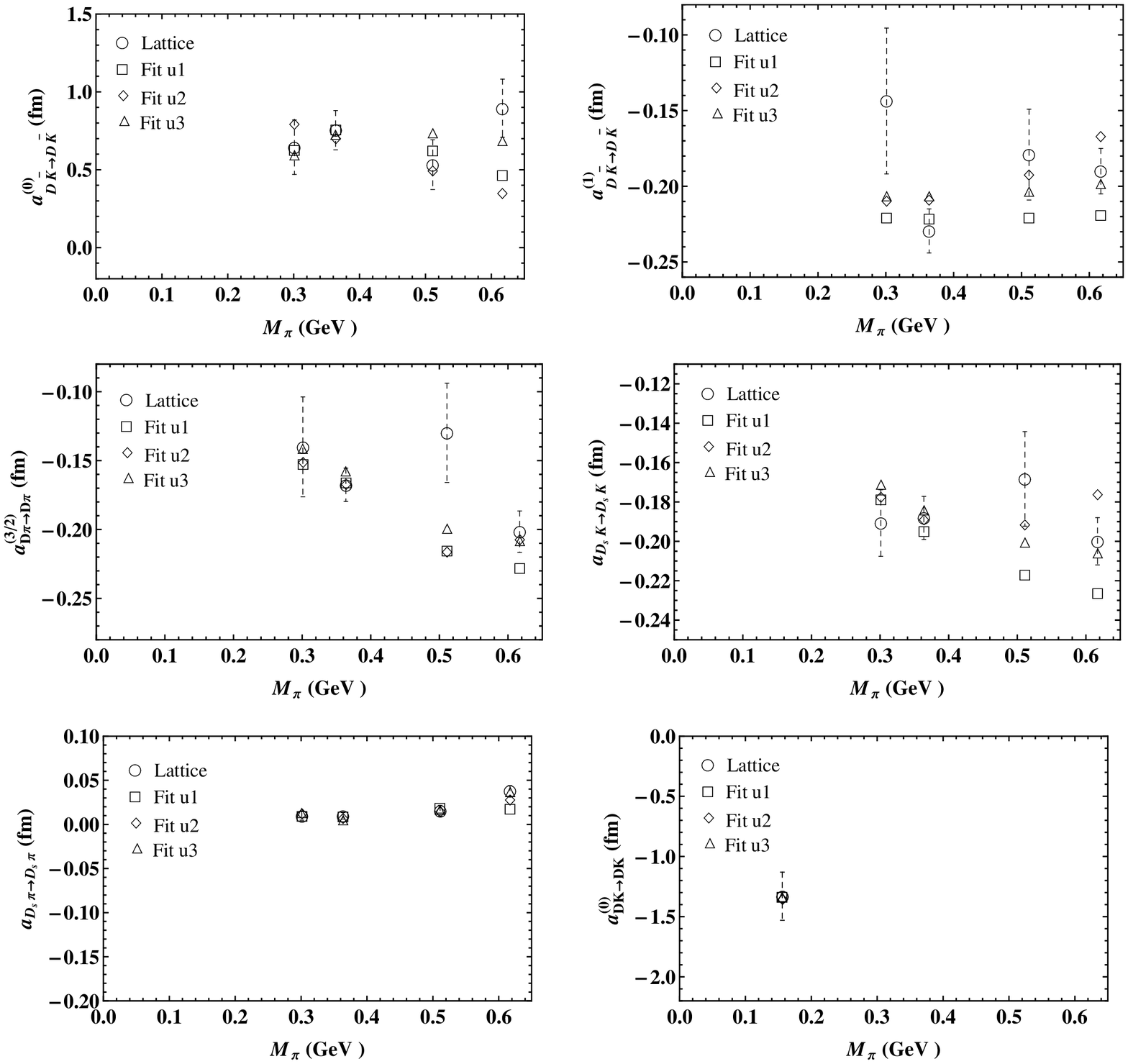}
\caption{\label{fig:fitu}Comparison of the results of the six-channel
fits through the iterated formula, Eq.~(\ref{scalenit}), to the
lattice data of the scattering lengths from
refs.~\cite{liu2013,mohl2013}.}
\end{figure}

The iterated scattering lengths formula is nonlinear which can make
the correlations become nonsense. However, the correlations with the
uncertainties of the constants can still be used to measure the
errors of the scattering lengths when the formula is linear in local
region. Fortunately, the three fits for the six channels at the
lattice points all fall in the approximate linear region, and there
are no unacceptable large values. The absolute value of the
correlation between $c_2$ and $c_4$ ($c_3$ and $c_5$) is still very
close to one, which also means that the non-linear effect is
marginal in the six channels at the lattice points. The values of
the LECs are also of natural size, and better results are obtained
in the three iterated fits. We present the corresponding scattering
lengths from the three iterated fits in Fig.~\ref{fig:fitu}. We can
clearly see that the results are improved significantly, especially
for $m_\pi\simeq 511, 617\,\text{MeV}$. In the channel $D_{s}\pi$
and $DK(I=0)$, the scattering lengths from all of the three iterated
fits are in good agreement with the lattice data points. However,
for the $D\bar{K}(I=1)$, the scattering length at $m_\pi \simeq 301\,\text{MeV}$
still has a large deviation because of the large error from the
lattice data point, but they have the same order of magnitude.

\begin{table*}[!t]
\centering
\resizebox{\textwidth}{!}{%
\begin{threeparttable}
\caption{\label{fittingresultu}Values of the various fits with the
correlations between the parameters through the iterated scattering
lengths formula, Eq.~(\ref{scalenit}). For a detailed description of
these fits, see the main text.}
\begin{tabular}{ccccccccccccccccccc}
\midrule \toprule
 & Fit u1 & $c_2$ & $c_3$ & $c_4$ & $c_5$ & $\bar{\kappa}$ & $\bar{e}_1$ & $\bar{e}_2$ & $\mu$& \\
\midrule
$c_2$ ($\text{GeV}^{-1}$)&$-4.60\pm 4.58$& $1.00$ & $-0.92$ & $-0.99$ & $0.89$ & $0.92$ & $0.70$ & $0.75$ & $0.19$ &\\
\midrule
$c_3$ ($\text{GeV}^{-1}$)&$3.32\pm 3.42$&  &$1.00$& $0.94$ & $-0.98$ & $-0.87$ & $-0.57$ & $-0.71$ & $-0.10$ &\\
\midrule
$c_4$ ($\text{GeV}^{-1}$)&$4.10\pm 3.47$&  &   & $1.00$ & $-0.91$ & $-0.92$ & $-0.68$  & $-0.75$ & $-0.17$ &\\
\midrule
$c_5$ ($\text{GeV}^{-1}$)&$-5.11\pm 3.73$&    &    & & $1.00$ & $0.92$ & $0.44$ & $0.78$ & $-0.07$ &\\
\midrule
$\bar{\kappa}$ ($\text{GeV}^{-2}$)&$-0.06\pm 0.17$&    &    & & & $1.00$ & $0.41$ &  $0.77$ & $-0.19$ &\\
\midrule
$\bar{e}_1$ ($\text{GeV}^{-3}$)&$2.38 \pm 1.65$&    &    & & & & $1.00$ & $0.39$ & $0.80$ &\\
\midrule
$\bar{e}_2$ ($\text{GeV}^{-3}$)&$-5.78 \pm 1.83$&    &     & & & & & $1.00$ & $-0.11$ &\\
\midrule
$\mu$ ($\text{GeV}$)&$1.03\pm 0.36$&    &     & & & & & & $1.00$ &\\
\midrule
$\chi^2/\text{d.o.f.}$&$\frac{3.90}{11-8}=1.30$&    &    &\\
\bottomrule
& Fit u2 & $c_2$ & $c_3$ & $c_4$ & $c_5$ & $\bar{\kappa}$ & $\bar{e}_1$ & $\bar{e}_2$ & $\mu$ & \\
\midrule
$c_2$ ($\text{GeV}^{-1}$)&$-3.03\pm 1.76$& $1.00$ &$-0.45$ & $-0.99$ & $0.48$ & $0.63$ & $0.35$ & $-0.46$ & $-0.53$ &\\
\midrule
$c_3$ ($\text{GeV}^{-1}$)&$0.96\pm 0.98$&  & $1.00$ & $0.52$ & $-0.98$ & $-0.63$ & $0.45$ & $0.54$ & $0.69$ &\\
\midrule
$c_4$ ($\text{GeV}^{-1}$)&$2.72\pm 1.27$&  &   & $1.00$ & $-0.55$ & $-0.66$ & $-0.28$ & $0.48$ & $0.56$ &\\
\midrule
$c_5$ ($\text{GeV}^{-1}$)&$-0.53\pm 1.23$&    &    & & $1.00$ & $0.73$ & $-0.46$ & $-0.61$ & $-0.78$ &\\
\midrule
$\bar{\kappa}$ ($\text{GeV}^{-2}$)&$0.48\pm 0.38$&    &    & & & $1.00$ & $-0.23$ & $-0.95$ & $-0.89$ &\\
\midrule
$\bar{e}_1$ ($\text{GeV}^{-3}$)&$1.29\pm 0.38$&    &    & & & & $1.00$ & $0.28$ & $0.41$ &\\
\midrule
$\bar{e}_2$ ($\text{GeV}^{-3}$)&$-8.19\pm 3.38$&    &     & & & & & $1.00$ & $0.81$ &\\
\midrule
$\mu$ ($\text{GeV}$)&$0.545\pm 0.124 $&    &     & & & & & & 1.00 &\\
\midrule
$\chi^2/\text{d.o.f.}$&$\frac{13.26}{16-8}=1.66$&    &    &\\
\bottomrule
& Fit u3 & $c_2$ & $c_3$ & $c_4$ & $c_5$ & $\bar{\kappa}$ & $\bar{e}_1$ & $\bar{e}_2$ & $\mu$ & \\
\midrule
$c_2$ ($\text{GeV}^{-1}$)&$-6.26\pm 1.36$& $1.00$ & $-0.83$ & $-0.99$ & $0.83$ & $0.90$ & $-0.83$ & $0.38$ & $-0.83$ &\\
\midrule
$c_3$ ($\text{GeV}^{-1}$)&$3.39\pm 1.84$&  & $1.00$ & $0.88$ & $-0.99$ & $-0.84$ & $0.97$ & $-0.41$ & $0.91$ & \\
\midrule
$c_4$ ($\text{GeV}^{-1}$)&$5.25 \pm 1.16$&  &   & $1.00$ & $-0.88$ & $-0.91$ & $0.88$ & $0.38$ & $0.86$ &\\
\midrule
$c_5$ ($\text{GeV}^{-1}$)&$-5.16 \pm 2.23$&    &    & & $1.00$ & $0.86$ & $-0.98$ & $0.54$ & $-0.95$ &\\
\midrule
$\bar{\kappa}$ ($\text{GeV}^{-2}$)&$-0.16\pm 0.06$&    &    & & & $1.00$ & $-0.84$ & $0.44$ & $-0.92$ &\\
\midrule
$\bar{e}_1$ ($\text{GeV}^{-3}$)&$2.18\pm 0.64$&    &    & & & & $1.00$ & $-0.52$ & $0.95$ &\\
\midrule
$\bar{e}_2$ ($\text{GeV}^{-3}$)&$-5.44 \pm 1.12$&    &     & & & & & $1.00$ & $-0.61$ &\\
\midrule
$\mu$ ($\text{GeV}$)& $1.164\pm 0.228$ &    &     & & & & & & $1.00$ &\\
\midrule
$\chi^2/\text{d.o.f.}$&$\frac{18.37}{21-8}=1.41$&    &    &\\
\bottomrule \midrule
\end{tabular}
\end{threeparttable}}%
\end{table*}

After the constants are determined using the lattice data points, we
make the predictions of the scattering lengths for the 11 channels
at the physical values in the iterated formula. The values of the
corresponding physical parameters are from PDG, see above. The
results from the three iterated fits are presented in
Tab.~\ref{dkscenlenu}. For comparison, the values from the other
two methods \cite{liu2013,guo2019} are listed in the table. The
errors of the scattering lengths are also estimated from the
statistical errors of the constants using the error propagation
formula with the correlations, as done in perturbative method. The
scattering lengths in our calculation have no imaginary parts
because the contributions include all the diagrams except the
iterated one-loop diagrams, and the iterated scattering length
formula only involves the real part.

For most channels in the three fits, reasonable values are obtained.
Unfortunately, the values from the $D\bar{K}(I=0)$ (Fit u1, u3),
$D_{s}\bar{K}$ (Fit u2) and $D\pi(I=1/2)$ (Fit u3), marked by
asterisks, are dominated by the nonlinear effect and have huge
uncertainties. The reason for the nonlinear effect is that a
bound-state pole can be generated by using parameters which are
close to the physical values. However, the bound-state pole is
physical only when all of the parameters are from the physical
values. The values with the asterisks are too large to be accepted
as the physical values. We can find that the values of the three
channels are not all in the nonlinear region for the three fits,
which indicates that the results are sensitive to the inputs. In
other words, we can not really improve the results through the
advanced analysis method. The more constraints, e.g., using more
lattice QCD data points, are one of the most effective methods to
obtain the physical scattering lengths. Except the values dominated
by the nonlinear effect, most values are either small negative or
positive numbers which indicate the small repulsive or attractive
interactions. However, at this time the values of the channel
$DK(I=0)$ for the three fits are large enough to roll over into
negative numbers which are attractive enough to generate the
well-known bound state $D_{s0}^{*}(2317)$. The result for this
channel is also consistent with the one from
refs.~\cite{liu2013,yao2015,guo2019} within errors. For the channels
$DK(I=1)$ and $D_{s}\bar{K}$, large negative numbers were obtained
in ref.~\cite{yao2015}, which does not mean that the bound state can
be generated because of the large imaginary parts.
\begin{table*}[!t]
\centering
\resizebox{\textwidth}{!}{%
\begin{threeparttable}
\caption{\label{dkscenlenu}Predictions of the scattering lengths
through the iterated formula, Eq.~(\ref{scalenit}). The LECs are
from the three different iterated fits, see
Tab.~\ref{fittingresultu}. The scattering lengths are in units of
fm. For the entries marked by asterisks, only the values without
uncertainties are given. See the main text for details.}
\begin{tabular}{ccccccccccccccccccc}
\midrule \toprule
& Fit u1 & Fit u2 & Fit u3 & Liu2013\cite{liu2013}& Guo2019\cite{guo2019} & \\
\midrule
$a_{DK}^{(1)}$ &$-0.03(4)$& $-0.01(3)$ & $-0.06(1)$ & $0.07^{+0.03}_{-0.03}+i0.17^{+0.02}_{-0.01}$ & $-0.01_{-0.03}^{+0.05}+i0.39^{+0.04}_{-0.04}$ &\\
\midrule
$a_{DK}^{(0)}$ &$-1.55(39)$& $-1.42(28)$ & $-1.67(45)$ & $-0.84^{+0.17}_{-0.22}$ & $-1.51^{+0.72}_{-2.35}$ &\\
\midrule
$a_{D\bar{K}}^{(1)}$ &$-0.23(2)$& $-0.24(2)$  & $-0.21(1)$ & $-0.20(1)$ & $-0.20^{+0.01}_{-0.01}$ &\\
\midrule
$a_{D\bar{K}}^{(0)}$ &$8.76^{\ast}$& $1.81(48)$   & $8.95^{\ast}$   & $0.84(15)$ & $21.9^{\ast}$  &\\
\midrule
$a_{D_s K}$ &$-0.14(3)$& $-0.17(3)$   & $-0.14(1)$  & $-0.18(1)$ & $-0.20^{+0.01}_{-0.01}$ &\\
\midrule
$a_{D_s \bar{K}}$ & $0.14(36)$ & $371.58^{\ast}$   & $0.05(18)$  & $-0.09^{+0.06}_{-0.05}+i0.44^{+0.05}_{-0.05}$ & $-0.57^{+0.06}_{-0.04}+i0.35^{+0.08}_{-0.07}$  & \\
\midrule
$a_{D\pi}^{(3/2)}$ & $-0.07(4)$ & $-0.06(2)$  &  $-0.05(1)$  & $-0.100(2)$  & $-0.103^{+0.003}_{-0.003}$ &\\
\midrule
$a_{D\pi}^{(1/2)}$ & $1.45(169)$ & $0.61(11)$   &  $6.00^{\ast}$  & $0.37^{+0.03}_{-0.02}$  & $0.40^{+0.03}_{-0.02}$ & \\
\midrule
$a_{D_s \pi}$ & $0.02(3)$ &  $0.03(2)$  &  $0.05(2)$  & $-0.002(1)$  & $0.012^{+0.003}_{-0.003}$ & \\
\midrule
$a_{D\eta}$ & $-0.04(14)$ & $0.03(8)$   &  $-0.09(2)$   &  & $0.29^{+0.15}_{-0.22}+i0.61^{+0.30}_{-0.26}$ &\\
\midrule
$a_{D_s \eta}$ & $-0.18(8)$ & $0.03(16)$  &  $-0.19(2)$ &  & $-0.39^{+0.05}_{-0.03}+i0.06^{+0.02}_{-0.02}$ &\\
\bottomrule \midrule
\end{tabular}
\end{threeparttable}}%
\end{table*}

Finally, we can estimate the meson and doubly charmed (bottom)
baryon scattering lengths through the heavy diquark-antiquark (HDA)
symmetry, as done in ref.~\cite{meng2019}. We take the values of the
physical baryon masses: $M_{\Xi_{cc}}=3621.55\,\text{MeV}$,
$M_{\Omega_{cc}}=3778\,\text{MeV}$,
$M_{\Xi_{bb}}=10202\,\text{MeV}$, and
$M_{{\Omega}_{bb}}=10359\,\text{MeV}$ from the PDG \cite{PDG2020}
and the relativistic quark model \cite{eber2002}. The other physical
parameters and LECs are taken as above. The perturbative and
iterated methods have also been used to predict the scattering
lengths. The results are presented in Appendixes~\ref{appA} (charmed
baryons) and \ref{appB} (bottom baryons). From
Tabs.~\ref{mdcbscenlenpfitp1}, \ref{mdcbscenlenpfitp2} and
\ref{mdcbscenlenpfitp3}, we can see clearly that the channel
$\bar{K}\Xi_{cc}(I=0)$ has strong attraction but still not strong
enough to generate any bound state. After using the iterated formula
in Tab.~\ref{mdcbscenlenu}, the value of the scattering length for
the channel $\bar{K}\Xi_{cc}(I=0)$ becomes large enough to roll over
into a negative number which is attractive enough to generate a
bound state. The same feature also exists in the channel
$\bar{K}\Xi_{bb}(I=0)$ from Tabs.~\ref{mdbbscenlenpfitp1},
\ref{mdbbscenlenpfitp2}, \ref{mdbbscenlenpfitp3} and
\ref{mdbbscenlenu}. The other features for the mesons and doubly
charmed (bottom) baryon scattering lengths are almost the same as
the above pseudoscalar meson and heavy meson scattering. For
comparison, we also list the meson and doubly charmed baryon
scattering lengths from ref.~\cite{guo2017} where the unitarized
chiral perturbation theory combined with the leading-order
amplitudes was used in the Tab.~\ref{mdcbscenlenu} of
ref.~\cite{guo2017}. The uncertainties of the scattering lengths for the channel
$K\Xi_{cc/bb}(I=0)$ also can become huge, which indicates that the
value is also dominated by the nonlinear effect.

\section{Summary}

In the past decades, many near-threshold non-conventional new hadron
states were discovered, many of which contain one, two, three or
even four heavy quarks. Unlike the nucleon-nucleon interaction which
has been studied very carefully for several decades, the
interactions either between the pseudoscalar mesons and heavy
hadrons or between two heavy hadrons are not known well. These
interactions are crucial to unveil the underlying structures of many
so-called hidden-charm tetraquarks, pentaquarks or even doubly
charmed $T_{cc}$ states.

Besides these manifestly exotic hadrons, the interpretation of the
conventional hadron spectrum also requires the precise knowledge of
the hadron-hadron interaction. For example, without the knowledge of
the $\pi\pi$, $\pi$K and KK interactions, we are unable to really
understand the low-lying scalar mesons below 1 GeV. Moreover, we all
know that there exist very strong couple-channel effects in the
formation of the low-lying parity-odd nucleon and hyperon
excitations such as $N(1535)$, $\Lambda (1405)$ etc. where the
pseudoscalar meson and nucleon/hyperon interaction is the underlying
driving force. Similarly, the strong $DK$ S-wave interaction leads
to the strong channel coupling between the $DK$ scattering channel
and the bare $c\bar s$ state in the quark model, which results in
the physical $D_{s0}^{*}(2317)$ state. Although there exist various
denotations or terminology such as the $DK$ molecule etc. for the
$D_{s0}^{*}(2317)$, the underlying dynamics is more or less the same
and the $DK$ interaction plays a pivotal role.

The scattering length provides valuable information of the hadron
interaction. In this work, we have calculated the threshold $T$
matrices of the light pseudoscalar meson and heavy meson scattering
to fourth order in heavy meson chiral perturbation theory. We
fitted the nonphysical lattice QCD data points to determine the LECs
through the perturbative and iterated methods, which led to a good
description of the scattering lengths at most of the lattice QCD
data points. The physical scattering lengths were obtained by
extrapolating the corresponding parameters to their physical values.
The values of the scattering lengths tend to be convergent at fourth
order for most of the channels in the perturbative method. The value
of the scattering length for the channel $DK(I=0)$, which involves
the well-known bound state $D_{s0}^{*}(2317)$, was obtained
correctly in the iterated method.

Based on heavy diquark-antiquark symmetry, we have also estimated
the meson and doubly charmed (bottom) baryon scattering lengths. We
want to emphasize that the bound states can be generated with high
probability in the channels $\bar{K}\Xi_{cc}(I=0)$ and
$\bar{K}\Xi_{bb}(I=0)$ through the kaon and $\Xi_{cc/bb}$
scattering. This doubly charmed molecule state is very similar to
the $D_{s0}^{*}(2317)$ state in many aspects because of the heavy
diquark-antiquark symmetry. The quantum number of this state is
$IJ^P=0{1\over 2}^-$.

Considering the orbital excitation is typically around 400-500 MeV
in quark model, this state may strongly couple with the P-wave
$\Omega_{cc/bb}$ state with the quark content $ccs/bbs$ and
$IJ^P=0{1\over 2}^-$. The physical state should lie slightly below
the $\bar{K}\Xi_{cc/bb}$ threshold. The only kinematically allowed
strong decay mode is $\Omega_{cc/bb} +\pi $, which violates the
isospin symmetry. Hence its strong decay width is around $\sim 0.1$
MeV or even less. Its E1 electromagnetic decay width is comparable
with its strong decay width, around 100 keV. We strongly urge LHCb Collaboration to look for the P-wave excitations of the
$\Omega_{cc/bb}$ and identify this very narrow signal in the coming
future.

\section*{Acknowledgments}
This work is supported by the National Natural Science Foundation of
China under Grants No. 11975033, No. 12070131001 and No. 12147127, and
China Postdoctoral Science Foundation (Grant No. 2021M700251).

\appendix\markboth{Appendix}{Appendix}
\renewcommand{\thesection}{\Alph{section}}
\numberwithin{equation}{section} \numberwithin{table}{section}

\clearpage
\section{Meson and doubly charmed baryon scattering lengths}
\label{appA}

\begin{table*}[!h]
\centering
\begin{threeparttable}
\caption{\label{mdcbscenlenpfitp1}Values of the mesons and doubly
charmed baryons scattering lengths in the calculation of the
perturbative formula, Eq.~(\ref{scalen}). The LECs are from the Fit
p1 of the Tab.~\ref{fittingresultp}. The scattering lengths are in
units of fm. Here $\Xi_{cc}=(\Xi_{cc}^{++},\Xi_{cc}^{+})^{T}$.}
\begin{tabular}{ccccccccccccccccccc}
\midrule \toprule
Fit p1 & $\mathcal{O}(p)$ & $\mathcal{O}(p^2)$ & $\mathcal{O}(p^3)$ & $\mathcal{O}(p^4)$ & \text{Total} & \\
\midrule
$a_{K\Xi_{cc}}^{(1)}$ &$-0.28$& $0.20$ &$-0.55$ &$0.37$& $-0.26(3)$ & \\
\midrule
$a_{K\Xi_{cc}}^{(0)}$ &$0.28$& $-0.04$ & $0.82$ & $-0.17$ & $0.89(7)$ & \\
\midrule
$a_{\bar{K}\Xi_{cc}}^{(1)}$ &$0.00$& $0.08$  & $-0.13+0.22i$  & $0.10-0.14i$ &$0.05(3)+0.08(18)i$&\\
\midrule
$a_{\bar{K}\Xi_{cc}}^{(0)}$ &$0.56$& $0.32$   & $0.86$   & $0.46$ & $2.20(103)$ &\\
\midrule
$a_{K\Omega_{cc}}$ &$0.28$& $0.20$   & $0.23+0.33i$   & $0.36+0.21i$ & $1.07(53)+0.54(24)i$ & \\
\midrule
$a_{\bar{K}\Omega_{cc}}$ & $-0.28$ & $0.20$   & $-0.42$   & $0.27$ & $-0.23(8)$ &  \\
\midrule
$a_{\pi\Xi_{cc}}^{(3/2)}$ & $-0.12$ & $0.03$  &  $-0.04$   & $0.04$ & $-0.10(4)$ &  \\
\midrule
$a_{\pi\Xi_{cc}}^{(1/2)}$ & $0.25$ & $0.03$   &  $0.02$   & $-0.01$  &  $0.28(2)$ &  \\
\midrule
$a_{\pi \Omega_{cc}}$ & $0.00$ &  $0.01$  &  $-0.04$  & $0.05$ & $0.02(2)$ &  \\
\midrule
$a_{\eta\Xi_{cc}}$ & $0.00$ & $0.13$   &  $0.0+0.24i$   & $0.02-0.11i$ & $0.15(9)+0.13(3)i$ & \\
\midrule
$a_{\eta\Omega_{cc}}$ & $0.00$ & $0.36$  &  $0.00+0.48i$   & $0.20+0.22i$ & $0.56(29)+0.70(14)i$ &  \\
\bottomrule \midrule
\end{tabular}
\end{threeparttable}
\end{table*}
~\\
~\\
~\\
~\\
\begin{table*}[!h]
\centering
\begin{threeparttable}
\caption{\label{mdcbscenlenpfitp2}The description is same with the
Tab.~\ref{mdcbscenlenpfitp1}, except that the LECs are from the the
Fit p2 of the Tab.~\ref{fittingresultp}.}
\begin{tabular}{ccccccccccccccccccc}
\midrule \toprule
Fit p2  & $\mathcal{O}(p)$ & $\mathcal{O}(p^2)$ & $\mathcal{O}(p^3)$ & $\mathcal{O}(p^4)$ & \text{Total} & \\
\midrule
$a_{K\Xi_{cc}}^{(1)}$ &$-0.28$& $0.24$ &$-0.61$ &$0.39$& $-0.25(3)$ & \\
\midrule
$a_{K\Xi_{cc}}^{(0)}$ &$0.28$& $-0.05$ & $0.87$ & $-0.26$ & $0.84(6)$ & \\
\midrule
$a_{\bar{K}\Xi_{cc}}^{(1)}$ &$0.00$& $0.09$  & $-0.13+0.22i$  & $0.07-0.18i$ &$0.03(3)+0.03(9)i$&\\
\midrule
$a_{\bar{K}\Xi_{cc}}^{(0)}$ &$0.56$& $0.39$   & $0.97$   & $0.52$ & $2.45(58)$ &\\
\midrule
$a_{K\Omega_{cc}}$ &$0.28$& $0.24$   & $0.29+0.33i$   & $0.37+0.28i$ & $1.19(29)+0.60(11)i$ & \\
\midrule
$a_{\bar{K}\Omega_{cc}}$ & $-0.28$ & $0.24$   & $-0.48$   & $0.27$ & $-0.25(4)$ &  \\
\midrule
$a_{\pi\Xi_{cc}}^{(3/2)}$ & $-0.12$ & $0.03$  &  $-0.04$   & $0.04$ & $-0.10(3)$ &  \\
\midrule
$a_{\pi\Xi_{cc}}^{(1/2)}$ & $0.25$ & $0.03$   &  $0.02$   & $-0.02$  &  $0.28(2)$ &  \\
\midrule
$a_{\pi \Omega_{cc}}$ & $0.00$ &  $0.01$  &  $-0.04$  & $0.05$ & $0.03(2)$ &  \\
\midrule
$a_{\eta\Xi_{cc}}$ & $0.00$ & $0.16$   &  $0.0+0.24i$   & $0.00-0.13i$ & $0.16(6)+0.11(1)i$ & \\
\midrule
$a_{\eta\Omega_{cc}}$ & $0.00$ & $0.42$  &  $0.00+0.48i$   & $0.22+0.26i$ & $0.64(17)+0.73(7)i$ &  \\
\bottomrule \midrule
\end{tabular}
\end{threeparttable}
\end{table*}

\begin{table*}[!h]
\centering
\begin{threeparttable}
\caption{\label{mdcbscenlenpfitp3}The description is same with the
Tab.~\ref{mdcbscenlenpfitp1}, except that the LECs are from the the
Fit p3 of the Tab.~\ref{fittingresultp}.}
\begin{tabular}{ccccccccccccccccccc}
\midrule \toprule
Fit p3 & $\mathcal{O}(p)$ & $\mathcal{O}(p^2)$ & $\mathcal{O}(p^3)$ & $\mathcal{O}(p^4)$ & \text{Total} &  \\
\midrule
$a_{K\Xi_{cc}}^{(1)}$ &$-0.28$& $0.13$ &$-0.37$ &$0.27$& $-0.25(2)$ & \\
\midrule
$a_{K\Xi_{cc}}^{(0)}$ &$0.28$& $0.04$ & $0.63$ & $-0.18$ & $0.78(5)$ & \\
\midrule
$a_{\bar{K}\Xi_{cc}}^{(1)}$ &$0.00$& $0.09$  & $-0.13+0.22i$  & $0.04-0.02i$ &$-0.00(2)+0.19(5)i$&\\
\midrule
$a_{\bar{K}\Xi_{cc}}^{(0)}$ &$0.56$& $0.17$   & $0.49$   & $0.38$ & $1.60(33)$ &\\
\midrule
$a_{K\Omega_{cc}}$ &$0.28$& $0.13$   & $0.04+0.33i$   & $0.28+0.03i$ & $0.73(16)+0.36(5)i$ & \\
\midrule
$a_{\bar{K}\Omega_{cc}}$ & $-0.28$ & $0.13$   & $-0.23$   & $0.22$ & $-0.17(3)$ &  \\
\midrule
$a_{\pi\Xi_{cc}}^{(3/2)}$ & $-0.12$ & $0.02$  &  $-0.03$   & $0.06$ & $-0.08(2)$ &  \\
\midrule
$a_{\pi\Xi_{cc}}^{(1/2)}$ & $0.25$ & $0.02$   &  $0.01$   & $0.02$  &  $0.29(1)$ &  \\
\midrule
$a_{\pi \Omega_{cc}}$ & $0.00$ &  $0.01$  &  $-0.04$  & $0.06$ & $0.03(1)$ &  \\
\midrule
$a_{\eta\Xi_{cc}}$ & $0.00$ & $0.10$   &  $0.0+0.24i$   & $0.11-0.06i$ & $0.21(3)+0.18(0)i$ & \\
\midrule
$a_{\eta\Omega_{cc}}$ & $0.00$ & $0.23$  &  $0.00+0.48i$   & $0.18+0.12i$ & $0.41(9)+0.60(4)i$ &  \\
\bottomrule \midrule
\end{tabular}
\end{threeparttable}
\end{table*}
~\\
~\\
~\\
~\\
~\\
\begin{table*}[!h]
\centering
\begin{threeparttable}
\caption{\label{mdcbscenlenu}Values of the mesons and doubly charmed
baryons scattering lengths in the calculation of the iterated
formula, Eq.~(\ref{scalenit}). The LECs are from the three different
iterated fits, see Tab.~\ref{fittingresultu}. The scattering lengths
are in units of fm. Here
$\Xi_{cc}=(\Xi_{cc}^{++},\Xi_{cc}^{+})^{T}$. For the entries marked
by asterisks, only the values without uncertainties are given. See
the main text for details.}
\begin{tabular}{ccccccccccccccccccc}
\midrule \toprule
& Fit u1 & Fit u2 & Fit u3 &  Guo2017 \cite{guo2017}& \\
\midrule
$a_{K\Xi_{cc}}^{(1)}$ &$-0.24(3)$& $-0.25(2)$ & $-0.22(2)$ & $-0.19^{+0.02}_{-0.02}$ & \\
\midrule
$a_{K\Xi_{cc}}^{(0)}$ &$-6.15^{\ast}$& $2.81^{\ast}$ & $-5.04^{\ast}$ & $5.2,-3.6,-1.4$ & \\
\midrule
$a_{\bar{K}\Xi_{cc}}^{(1)}$ &$-0.04(4)$& $-0.01(3)$  & $-0.06(2)$ & $-0.22^{+0.14}_{-0.14}+i0.45^{+0.00}_{-0.09}$   &\\
\midrule
$a_{\bar{K}\Xi_{cc}}^{(0)}$ &$-1.19(21)$& $-1.29(24)$   & $-1.20(19)$   & $-0.49^{+0.10}_{-0.19}$ &\\
\midrule
$a_{K\Omega_{cc}}$ &$0.16(43)$& $-7.80^{\ast}$   & $0.06(20)$  & $-0.55^{+0.11}_{-0.16}+i0.13^{+0.19}_{-0.07}$ &\\
\midrule
$a_{\bar{K}\Omega_{cc}}$ & $-0.15(3)$ & $-0.18(3)$   & $-0.15(1)$  & $-0.19^{+0.02}_{-0.02}$ &\\
\midrule
$a_{\pi\Xi_{cc}}^{(3/2)}$ & $-0.07(3)$ & $-0.07(2)$  &  $-0.05(1)$  & $-0.095^{+0.003}_{-0.004}$ & \\
\midrule
$a_{\pi\Xi_{cc}}^{(1/2)}$ & $1.73^{\ast}$ & $0.65(12)$   &  $16.26^{\ast}$  & $0.55^{+0.16}_{-0.10}$ & \\
\midrule
$a_{\pi\Omega_{cc}}$ & $0.02(3)$ &  $0.03(2)$  &  $0.05(1)$  &  $0.03^{+0.01}_{-0.01}$ & \\
\midrule
$a_{\eta \Xi_{cc}}$ & $-0.05(15)$ & $0.03(10)$   &  $-0.09(2)$   &  $-0.72^{+0.21}_{-0.17}+i0.30^{+1.10}_{-0.18}$\\
\midrule
$a_{\eta\Omega_{cc}}$ & $-0.19(8)$ & $0.04(18)$  &  $-0.20(2)$ & $-0.26^{+0.03}_{-0.03}+i0.02^{+0.02}_{-0.01}$ &\\
\bottomrule \midrule
\end{tabular}
\end{threeparttable}
\end{table*}

\clearpage
\section{Meson and doubly bottom baryon scattering lengths}
\label{appB}

\begin{table*}[!h]
\centering
\begin{threeparttable}
\caption{\label{mdbbscenlenpfitp1}Values of the mesons and doubly
bottom baryons scattering lengths in the calculation of the
perturbative formula, Eq.~(\ref{scalen}). The LECs are from the Fit
p1 of the Tab.~\ref{fittingresultp}. The scattering lengths are in
units of fm. Here $\Xi_{bb}=(\Xi_{bb}^{0},\Xi_{bb}^{-})^{T}$.}
\begin{tabular}{ccccccccccccccccccc}
\midrule \toprule
Fit p1 & $\mathcal{O}(p)$ & $\mathcal{O}(p^2)$ & $\mathcal{O}(p^3)$ & $\mathcal{O}(p^4)$ & \text{Total} & \\
\midrule
$a_{K\Xi_{bb}}^{(1)}$ &$-0.31$& $0.22$ &$-0.60$ &$0.40$& $-0.28(3)$ & \\
\midrule
$a_{K\Xi_{bb}}^{(0)}$ &$0.31$& $-0.04$ & $0.88$ & $-0.19$ & $0.96(8)$ & \\
\midrule
$a_{\bar{K}\Xi_{bb}}^{(1)}$ &$0.00$& $0.09$  & $-0.15+0.23i$  & $0.11-0.15i$ &$0.05(4)+0.08(19)i$&\\
\midrule
$a_{\bar{K}\Xi_{bb}}^{(0)}$ &$0.61$& $0.35$   & $0.93$   & $0.50$ & $2.39(112)$ &\\
\midrule
$a_{K\Omega_{bb}}$ &$0.31$& $0.22$   & $0.25+0.35i$   & $0.39+0.23i$ & $1.16(57)+0.58(26)i$ & \\
\midrule
$a_{\bar{K}\Omega_{bb}}$ & $-0.31$ & $0.22$   & $-0.45$   & $0.29$ & $-0.25(9)$ &  \\
\midrule
$a_{\pi\Xi_{bb}}^{(3/2)}$ & $-0.13$ & $0.03$  &  $-0.04$   & $0.04$ & $-0.10(4)$ &  \\
\midrule
$a_{\pi\Xi_{bb}}^{(1/2)}$ & $0.26$ & $0.03$   &  $0.02$   & $-0.02$  &  $0.28(2)$ &  \\
\midrule
$a_{\pi \Omega_{bb}}$ & $0.00$ &  $0.01$  &  $-0.04$  & $0.05$ & $0.02(2)$ &  \\
\midrule
$a_{\eta\Xi_{bb}}$ & $0.00$ & $0.14$   &  $0.0+0.26i$   & $0.02-0.12i$ & $0.16(10)+0.14(3)i$ & \\
\midrule
$a_{\eta\Omega_{bb}}$ & $0.00$ & $0.39$  &  $0.00+0.52i$   & $0.22+0.24i$ & $0.61(32)+0.76(16)i$ &  \\
\bottomrule \midrule
\end{tabular}
\end{threeparttable}
\end{table*}
~\\
~\\
\begin{table*}[!h]
\centering
\begin{threeparttable}
\caption{\label{mdbbscenlenpfitp2}The description is same with the
Tab.~\ref{mdbbscenlenpfitp1}, except that the LECs are from the the
Fit p2 of the Tab.~\ref{fittingresultp}.}
\begin{tabular}{ccccccccccccccccccc}
\midrule \toprule
Fit p2 & $\mathcal{O}(p)$ & $\mathcal{O}(p^2)$ & $\mathcal{O}(p^3)$ & $\mathcal{O}(p^4)$ & \text{Total} & \\
\midrule
$a_{K\Xi_{bb}}^{(1)}$ &$-0.31$& $0.26$ &$-0.66$ &$0.43$& $-0.27(3)$ & \\
\midrule
$a_{K\Xi_{bb}}^{(0)}$ &$0.31$& $-0.06$ & $0.95$ & $-0.28$ & $0.91(7)$ & \\
\midrule
$a_{\bar{K}\Xi_{bb}}^{(1)}$ &$0.00$& $0.10$  & $-0.14+0.23i$  & $0.07-0.20i$ &$0.03(3)+0.04(10)i$&\\
\midrule
$a_{\bar{K}\Xi_{bb}}^{(0)}$ &$0.61$& $0.42$   & $1.05$   & $0.57$ & $2.65(63)$ &\\
\midrule
$a_{K\Omega_{bb}}$ &$0.31$& $0.26$   & $0.31+0.35i$   & $0.40+0.30i$ & $1.28(32)+0.65(12)i$ & \\
\midrule
$a_{\bar{K}\Omega_{bb}}$ & $-0.31$ & $0.26$   & $-0.51$   & $0.29$ & $-0.27(5)$ &  \\
\midrule
$a_{\pi\Xi_{bb}}^{(3/2)}$ & $-0.13$ & $0.03$  &  $-0.04$   & $0.04$ & $-0.10(3)$ &  \\
\midrule
$a_{\pi\Xi_{bb}}^{(1/2)}$ & $0.26$ & $0.03$   &  $0.02$   & $-0.02$  &  $0.29(2)$ &  \\
\midrule
$a_{\pi \Omega_{bb}}$ & $0.00$ &  $0.01$  &  $-0.04$  & $0.05$ & $0.03(2)$ &  \\
\midrule
$a_{\eta\Xi_{bb}}$ & $0.00$ & $0.17$   &  $0.0+0.26i$   & $0.00-0.14i$ & $0.17(6)+0.12(1)i$ & \\
\midrule
$a_{\eta\Omega_{bb}}$ & $0.00$ & $0.46$  &  $0.00+0.52i$   & $0.24+0.28i$ & $0.70(18)+0.80(8)i$ &  \\
\bottomrule \midrule
\end{tabular}
\end{threeparttable}
\end{table*}

\begin{table*}[!h]
\centering
\begin{threeparttable}
\caption{\label{mdbbscenlenpfitp3}The description is same with the
Tab.~\ref{mdbbscenlenpfitp1}, except that the LECs are from the the
Fit p3 of the Tab.~\ref{fittingresultp}.}
\begin{tabular}{ccccccccccccccccccc}
\midrule \toprule
Fit p3 & $\mathcal{O}(p)$ & $\mathcal{O}(p^2)$ & $\mathcal{O}(p^3)$ & $\mathcal{O}(p^4)$ & \text{Total} & \\
\midrule
$a_{K\Xi_{bb}}^{(1)}$ &$-0.31$& $0.14$ &$-0.40$ & $0.29$ & $-0.27(2)$ & \\
\midrule
$a_{K\Xi_{bb}}^{(0)}$ &$0.31$& $0.05$ & $0.69$ & $-0.19$ & $0.84(5)$ & \\
\midrule
$a_{\bar{K}\Xi_{bb}}^{(1)}$ &$0.00$& $0.09$  & $-0.14+0.23i$  & $0.05-0.02i$ &$-0.00(2)+0.21(5)i$&\\
\midrule
$a_{\bar{K}\Xi_{bb}}^{(0)}$ &$0.61$& $0.19$   & $0.53$   & $0.41$ & $1.74(36)$ &\\
\midrule
$a_{K\Omega_{bb}}$ &$0.31$& $0.14$   & $0.05+0.35i$   & $0.30+0.04i$ & $0.80(18)+0.39(5)i$ & \\
\midrule
$a_{\bar{K}\Omega_{bb}}$ & $-0.31$ & $0.14$   & $-0.25$   & $0.24$ & $-0.18(3)$ &  \\
\midrule
$a_{\pi\Xi_{bb}}^{(3/2)}$ & $-0.13$ & $0.02$  &  $-0.03$   & $0.07$ & $-0.08(2)$ &  \\
\midrule
$a_{\pi\Xi_{bb}}^{(1/2)}$ & $0.26$ & $0.02$   &  $0.01$   & $0.02$  &  $0.29(1)$ &  \\
\midrule
$a_{\pi \Omega_{bb}}$ & $0.00$ &  $0.01$  &  $-0.04$  & $0.06$ & $0.03(1)$ &  \\
\midrule
$a_{\eta\Xi_{bb}}$ & $0.00$ & $0.11$   &  $0.0+0.26i$   & $0.12-0.07i$ & $0.23(3)+0.19(0)i$ & \\
\midrule
$a_{\eta\Omega_{bb}}$ & $0.00$ & $0.25$  &  $0.00+0.52i$   & $0.19+0.13i$ & $0.45(9)+0.65(4)i$ &  \\
\bottomrule \midrule
\end{tabular}
\end{threeparttable}
\end{table*}
~\\
~\\
~\\
~\\
~\\
~\\
~\\
\begin{table*}[!h]
\centering
\begin{threeparttable}
\caption{\label{mdbbscenlenu} Values of the mesons and doubly bottom
baryons scattering lengths in the calculation of the iterated
formula, Eq.~(\ref{scalenit}). The LECs are from the three different
iterated fits, see Tab.~\ref{fittingresultu}. The scattering lengths
are in units of fm. Here $\Xi_{bb}=(\Xi_{bb}^{0},\Xi_{bb}^{-})^{T}$.
For the entries marked by asterisks, only the values without
uncertainties are given. See the main text for details.}
\begin{tabular}{ccccccccccccccccccc}
\midrule \toprule
& Fit u1 & Fit u2 & Fit u3 & \\
\midrule
$a_{K\Xi_{bb}}^{(1)}$ &$-0.24(3)$& $-0.27(2)$ & $-0.23(2)$ & \\
\midrule
$a_{K\Xi_{bb}}^{(0)}$ &$-2.84^{\ast}$& $4.51^{\ast}$ & $-2.43^{\ast}$ & \\
\midrule
$a_{\bar{K}\Xi_{bb}}^{(1)}$ &$-0.04(4)$& $-0.01(4)$  & $-0.07(2)$ &\\
\midrule
$a_{\bar{K}\Xi_{bb}}^{(0)}$ &$-1.02(18)$& $-1.22(22)$   & $-1.00(23)$   &\\
\midrule
$a_{K\Omega_{bb}}$ &$0.18(49)$& $-4.52^{\ast}$   & $0.06(22)$  & \\
\midrule
$a_{\bar{K}\Omega_{bb}}$ & $-0.16(3)$ & $-0.19(3)$   & $-0.16(1)$  &  \\
\midrule
$a_{\pi\Xi_{bb}}^{(3/2)}$ & $-0.07(3)$ & $-0.07(2)$  &  $-0.05(1)$  &  \\
\midrule
$a_{\pi\Xi_{bb}}^{(1/2)}$ & $1.99^{\ast}$ & $0.68(13)$   &  $2.01^{\ast}$  &  \\
\midrule
$a_{\pi\Omega_{bb}}$ & $0.02(4)$ &  $0.04(2)$  &  $0.05(2)$  &  \\
\midrule
$a_{\eta \Xi_{bb}}$ & $-0.05(16)$ & $0.04(11)$   &  $-0.10(2)$   & \\
\midrule
$a_{\eta\Omega_{bb}}$ & $-0.20(8)$ & $0.04(20)$  &  $-0.20(2)$ &  \\
\bottomrule \midrule
\end{tabular}
\end{threeparttable}
\end{table*}

\bibliographystyle{unsrt}
\bibliography{latextemplate}

\providecommand{\noopsort}[1]{}\providecommand{\singleletter}[1]{#1}%
\begin{thebibliography}{10}

\bibitem{baba2003}
B.~Aubert et~al.
\newblock {Observation of a narrow meson decaying to $D_s^+\pi^0$ at a mass of
  2.32-GeV/c$^2$}.
\newblock {\em Phys. Rev. Lett.}, 90:242001, 2003.

\bibitem{bell2003}
P.~Krokovny et~al.
\newblock {Observation of the ${D}_{sJ}(2317)$ and ${D}_{sJ}(2457)$ in $B$
  Decays}.
\newblock {\em Phys. Rev. Lett.}, 91:262002, 2003.

\bibitem{cleo2003}
D.~Besson et~al.
\newblock {Observation of a narrow resonance of mass $2.46$ GeV${/c}^{2}$
  decaying to ${D}_{s}^{*+}{\ensuremath{\pi}}^{0}$ and confirmation of the
  ${D}_{\mathrm{sJ}}^{*}(2317)$ state}.
\newblock {\em Phys. Rev. D}, 68:032002, 2003.
\newblock [Erratum: Phys.Rev.D 75, 119908 (2007)].

\bibitem{godf1985}
S.~Godfrey and Nathan Isgur.
\newblock {Mesons in a Relativized Quark Model with Chromodynamics}.
\newblock {\em Phys. Rev. D}, 32:189--231, 1985.

\bibitem{godf1991}
Stephen Godfrey and Richard Kokoski.
\newblock Properties of $p$-wave mesons with one heavy quark.
\newblock {\em Phys. Rev. D}, 43:1679--1687, 1991.

\bibitem{dipi2001}
M.~Di~Pierro and E.~Eichten.
\newblock Excited heavy-light systems and hadronic transitions.
\newblock {\em Phys. Rev. D}, 64:114004, 2001.

\bibitem{dai1994}
Yuan-Ben Dai, Chao-Shang Huang, and Hong-Ying Jin.
\newblock Heavy meson spectra from relativistic {B}-{S} equations to the order
  1/{M}.
\newblock {\em Phys. Lett. B}, 331:174--178, 1994.

\bibitem{barn2003}
T.~Barnes, F.~E. Close, and H.~J. Lipkin.
\newblock Implications of a {DK} molecule at 2.32 {G}e{V}.
\newblock {\em Phys. Rev. D}, 68:054006, 2003.

\bibitem{chen2004}
Yu-Qi Chen and Xue-Qian Li.
\newblock Comprehensive {F}our-{Q}uark {I}nterpretation of ${D}_{s}(2317)$,
  ${D}_{s}(2457)$, and ${D}_{s}(2632)$.
\newblock {\em Phys. Rev. Lett.}, 93:232001, 2004.

\bibitem{guo2006}
Feng-Kun Guo, Peng-Nian Shen, Huan-Ching Chiang, Rong-Gang Ping, and Bing-Song
  Zou.
\newblock {Dynamically generated 0+ heavy mesons in a heavy chiral unitary
  approach}.
\newblock {\em Phys. Lett. B}, 641:278--285, 2006.

\bibitem{guo2007}
Feng-Kun Guo, Peng-Nian Shen, and Huan-Ching Chiang.
\newblock Dynamically generated 1+ heavy mesons.
\newblock {\em Phys. Lett. B}, 647(2):133--139, 2007.

\bibitem{chen2003}
Hai-Yang Cheng and Wei-Shu Hou.
\newblock B decays as spectroscope for charmed four-quark states.
\newblock {\em Phys. Lett. B}, 566(3):193--200, 2003.

\bibitem{brow2004}
Thomas~E Browder, Sandip Pakvasa, and Alexey~A Petrov.
\newblock Comment on the new ${D}_{s}^{(\ast)+}\pi^{0}$ resonances.
\newblock {\em Phys. Lett. B}, 578(3):365--368, 2004.

\bibitem{dmit2005}
V.~Dmitra\ifmmode \check{s}\else \v{s}\fi{}inovi\ifmmode~\acute{c}\else
  \'{c}\fi{}.
\newblock ${D}_{s0}^{+}(2317)$-${D}_{0}(2308)$ mass difference as evidence for
  tetraquarks.
\newblock {\em Phys. Rev. Lett.}, 94:162002, 2005.

\bibitem{brac2005}
M.E. Bracco, A.~Lozea, R.D. Matheus, F.S. Navarra, and M.~Nielsen.
\newblock Disentangling two- and four-quark state pictures of the charmed
  scalar mesons.
\newblock {\em Phys. Lett. B}, 624(3):217--222, 2005.

\bibitem{beve2003}
Eef van Beveren and George Rupp.
\newblock Observed ${D}_{s}(2317)$ and tentative $d(2100--2300)$ as the charmed
  cousins of the light scalar nonet.
\newblock {\em Phys. Rev. Lett.}, 91:012003, 2003.

\bibitem{bali2003}
Gunnar~S. Bali.
\newblock ${D}_{\mathrm{sj}}^{+}(2317):$ what can the lattice say?
\newblock {\em Phys. Rev. D}, 68:071501, 2003.

\bibitem{doug2003}
A~Dougall, R.D Kenway, C.M Maynard, and C~McNeile.
\newblock The spectrum of ds mesons from lattice qcd.
\newblock {\em Physics Letters B}, 569(1):41--44, 2003.

\bibitem{flyn2007}
Jonathan~M Flynn and Juan Nieves.
\newblock {Elastic s-wave $B\pi$, $D\pi$, $D K$ and $K \pi$ scattering from
  lattice calculations of scalar form-factors in semileptonic decays}.
\newblock {\em Phys. Rev. D}, 75:074024, 2007.

\bibitem{liu2013}
Liuming Liu, Kostas Orginos, Feng-Kun Guo, Christoph Hanhart, and Ulf-G.
  Mei\ss{}ner.
\newblock Interactions of charmed mesons with light pseudoscalar mesons from
  lattice {QCD} and implications on the nature of the ${D}_{s0}^{*}(2317)$.
\newblock {\em Phys. Rev. D}, 87:014508, 2013.

\bibitem{lang2014}
C.~B. Lang, Luka Leskovec, Daniel Mohler, Sasa Prelovsek, and R.~M. Woloshyn.
\newblock {$D_s$ mesons with $DK$ and $D^{\ast}K$ scattering near threshold}.
\newblock {\em Phys. Rev. D}, 90(3):034510, 2014.

\bibitem{mohl2013}
Daniel Mohler, C.~B. Lang, Luka Leskovec, Sasa Prelovsek, and R.~M. Woloshyn.
\newblock ${D}_{s0}^{*}\mathbf{(}2317\mathbf{)}$ meson and {D}-meson-kaon
  scattering from lattice {QCD}.
\newblock {\em Phys. Rev. Lett.}, 111:222001, 2013.

\bibitem{chen2017}
Hua-Xing Chen, Wei Chen, Xiang Liu, Yan-Rui Liu, and Shi-Lin Zhu.
\newblock {A review of the open charm and open bottom systems}.
\newblock {\em Rept. Prog. Phys.}, 80(7):076201, 2017.

\bibitem{walk2009}
A.~Walker-Loud, H.-W. Lin, D.~G. Richards, R.~G. Edwards, M.~Engelhardt, G.~T.
  Fleming, Ph. H\"agler, B.~Musch, M.~F. Lin, H.~Meyer, J.~W. Negele, A.~V.
  Pochinsky, M.~Procura, S.~Syritsyn, C.~J. Morningstar, K.~Orginos, D.~B.
  Renner, and W.~Schroers.
\newblock Light hadron spectroscopy using domain wall valence quarks on an
  asqtad sea.
\newblock {\em Phys. Rev. D}, 79:054502, 2009.

\bibitem{bali2017}
Gunnar~S. Bali, Sara Collins, Antonio Cox, and Andreas Sch\"afer.
\newblock {Masses and decay constants of the $D_{s0}^*(2317)$ and
  $D_{s1}(2460)$ from $N_f=2$ lattice QCD close to the physical point}.
\newblock {\em Phys. Rev. D}, 96(7):074501, 2017.

\bibitem{alex2019}
Constantia Alexandrou, Joshua Berlin, Jacob Finkenrath, Theodoros Leontiou, and
  Marc Wagner.
\newblock {Tetraquark interpolating fields in a lattice QCD investigation of
  the $D_{s0}^\ast(2317)$ meson}.
\newblock {\em Phys. Rev. D}, 101(3):034502, 2020.

\bibitem{wein1990}
Steven Weinberg.
\newblock {Nuclear forces from chiral Lagrangians}.
\newblock {\em Phys. Lett. B}, 251:288--292, 1990.

\bibitem{wein1991}
Steven Weinberg.
\newblock {Effective chiral Lagrangians for nucleon - pion interactions and
  nuclear forces}.
\newblock {\em Nucl. Phys. B}, 363:3--18, 1991.

\bibitem{gass1987}
J.~Gasser, M.~E. Sainio, and A.~Svarc.
\newblock {Nucleons with Chiral Loops}.
\newblock {\em Nucl. Phys. B}, 307:779--853, 1988.

\bibitem{jenk1991}
E.~E. Jenkins and A.~V. Manohar.
\newblock {Baryon chiral perturbation theory using a heavy fermion Lagrangian}.
\newblock {\em Phys. Lett. B}, 255:558--562, 1991.

\bibitem{bern1992}
V.~Bernard, N.~Kaiser, J.~Kambor, and U.-G. Mei{\ss}ner.
\newblock {Chiral structure of the nucleon}.
\newblock {\em Nucl. Phys. B}, 388:315--345, 1992.

\bibitem{ordo1992}
C.~Ordóñez and U.~{van Kolck}.
\newblock Chiral lagrangians and nuclear forces.
\newblock {\em Phys. Lett. B}, 291(4):459--464, 1992.

\bibitem{epel1998}
E.~Epelbaoum, W.~Glöckle, and Ulf-G. Meißner.
\newblock Nuclear forces from chiral lagrangians using the method of unitary
  transformation (i): Formalism.
\newblock {\em Nucl. Phys. A}, 637(1):107--134, 1998.

\bibitem{fett2000}
N.~Fettes and U.-G. Mei{\ss}ner.
\newblock {Pion nucleon scattering in chiral perturbation theory (II): Fourth
  order calculation}.
\newblock {\em Nucl. Phys. A}, 676:311, 2000.

\bibitem{kais19971}
N.~Kaiser, R.~Brockmann, and W.~Weise.
\newblock {Peripheral nucleon-nucleon phase shifts and chiral symmetry}.
\newblock {\em Nucl. Phys. A}, 625:758--788, 1997.

\bibitem{mach2011}
R.~Machleidt and D.~R. Entem.
\newblock {Chiral effective field theory and nuclear forces}.
\newblock {\em Phys. Rept.}, 503:1--75, 2011.

\bibitem{kang2013}
Xian-Wei Kang, Johann Haidenbauer, and Ulf-G. Mei\ss{}ner.
\newblock {Antinucleon-nucleon interaction in chiral effective field theory}.
\newblock {\em JHEP}, 02:113, 2014.

\bibitem{ente2015}
D.~R. Entem, N.~Kaiser, R.~Machleidt, and Y.~Nosyk.
\newblock {Peripheral nucleon-nucleon scattering at fifth order of chiral
  perturbation theory}.
\newblock {\em Phys. Rev. C}, 91(1):014002, 2015.

\bibitem{kais2020}
N.~Kaiser.
\newblock {Density-dependent NN interaction from subsubleading chiral 3N
  forces: Intermediate-range contributions}.
\newblock {\em Phys. Rev. C}, 101(1):014001, 2020.

\bibitem{kais2001}
N.~Kaiser.
\newblock {Chiral corrections to kaon nucleon scattering lengths}.
\newblock {\em Phys. Rev. C}, 64:045204, 2001.
\newblock [Erratum: Phys. Rev.C73,069902(2006)].

\bibitem{liu20071}
Y.-R. Liu and S.-L. Zhu.
\newblock {Meson-baryon scattering lengths in HB$\chi$PT}.
\newblock {\em Phys. Rev. D}, 75:034003, 2007.

\bibitem{haid2013}
J.~Haidenbauer, S.~Petschauer, N.~Kaiser, U.-G. Meißner, A.~Nogga, and
  W.~Weise.
\newblock Hyperon–nucleon interaction at next-to-leading order in chiral
  effective field theory.
\newblock {\em Nucl. Phys. A}, 915:24--58, 2013.

\bibitem{huan2015}
B.-L. Huang and Y.-D. Li.
\newblock {Kaon-nucleon scattering to one-loop order in heavy baryon chiral
  perturbation theory}.
\newblock {\em Phys. Rev. D}, 92(11):114033, 2015.
\newblock [Erratum: Phys. Rev.D95,019903(2017)].

\bibitem{huan2017}
B.-L. Huang, J.-S. Zhang, Y.-D. Li, and N.~Kaiser.
\newblock {Meson-baryon scattering to one-loop order in heavy baryon chiral
  perturbation theory}.
\newblock {\em Phys. Rev. D}, 96(11):016021, 2017.

\bibitem{huan20201}
B.-L. Huang and J.~Ou-Yang.
\newblock {Pion-nucleon scattering to $\mathcal{O}(p^3)$ in heavy baryon SU(3)
  chiral perturbation theory}.
\newblock {\em Phys. Rev. D}, 101:056021, 2020.

\bibitem{huan20202}
B.-L. Huang.
\newblock {Pion-nucleon scattering to order $p^4$ in SU(3) heavy baryon chiral
  perturbation theory}.
\newblock {\em Phys. Rev. D}, 102:116001, 2020.

\bibitem{huan2021}
B.-L. Huang, J.-B. Cheng, and S.-L. Zhu.
\newblock {Peripheral nucleon-nucleon scattering at next-to-next-to-leading
  order in SU(3) heavy baryon chiral perturbation theory}.
\newblock {\em Phys. Rev. D}, 104:116030, 2021.

\bibitem{wise1992}
Mark~B. Wise.
\newblock {Chiral perturbation theory for hadrons containing a heavy quark}.
\newblock {\em Phys. Rev. D}, 45(7):R2188, 1992.

\bibitem{bern2008}
Véronique Bernard.
\newblock Chiral perturbation theory and baryon properties.
\newblock {\em Progress in Particle and Nuclear Physics}, 60(1):82--160, 2008.

\bibitem{liu2009}
Yan-Rui Liu, Xiang Liu, and Shi-Lin Zhu.
\newblock {Light Pseudoscalar Meson and Heavy Meson Scattering Lengths}.
\newblock {\em Phys. Rev. D}, 79:094026, 2009.

\bibitem{guo2009}
Feng-Kun Guo, Christoph Hanhart, and Ulf-G. Meissner.
\newblock {Interactions between heavy mesons and Goldstone bosons from chiral
  dynamics}.
\newblock {\em Eur. Phys. J. A}, 40:171--179, 2009.

\bibitem{geng2010}
L.~S. Geng, N.~Kaiser, J.~Martin-Camalich, and W.~Weise.
\newblock Low-energy interactions of nambu-goldstone bosons with $d$ mesons in
  covariant chiral perturbation theory.
\newblock {\em Phys. Rev. D}, 82:054022, 2010.

\bibitem{wang2012}
P.~Wang and X.~G. Wang.
\newblock Publisher's note: Study of ${0}^{\mathbf{+}}$ states with open charm
  in the unitarized heavy meson chiral approach [phys. rev. d 86, 014030
  (2012)].
\newblock {\em Phys. Rev. D}, 86:039903, Aug 2012.

\bibitem{alte2014}
M.~Altenbuchinger, L.~S. Geng, and W.~Weise.
\newblock {Scattering lengths of Nambu-Goldstone bosons off $D$ mesons and
  dynamically generated heavy-light mesons}.
\newblock {\em Phys. Rev. D}, 89(1):014026, 2014.

\bibitem{yao2015}
De-Liang Yao, Meng-Lin Du, Feng-Kun Guo, and Ulf-G. Mei\ss{}ner.
\newblock {One-loop analysis of the interactions between charmed mesons and
  Goldstone bosons}.
\newblock {\em JHEP}, 11:058, 2015.

\bibitem{guo2019}
Zhi-Hui Guo, Liuming Liu, Ulf-G Mei\ss{}ner, J.~A. Oller, and A.~Rusetsky.
\newblock {Towards a precise determination of the scattering amplitudes of the
  charmed and light-flavor pseudoscalar mesons}.
\newblock {\em Eur. Phys. J. C}, 79(1):13, 2019.

\bibitem{bora1997}
B.~Borasoy and U.-G. Mei{\ss}ner.
\newblock {Chiral expansion of baryon masses and sigma-terms}.
\newblock {\em Annals Phys.}, 254:192--232, 1997.

\bibitem{kais1995}
N.~Kaiser, P.~B. Siegel, and W.~Weise.
\newblock {Chiral dynamics and the low-energy kaon-nucleon interaction}.
\newblock {\em Nucl. Phys. A}, 594:325--345, 1995.

\bibitem{hofm2004}
J.~Hofmann and M.~F.~M. Lutz.
\newblock {Open charm meson resonances with negative strangeness}.
\newblock {\em Nucl. Phys. A}, 733:142--152, 2004.

\bibitem{doba2014}
J.~Dobaczewski, W.~Nazarewicz, and P.-G. Reinhard.
\newblock {Error estimates of theoretical models: a Guide}.
\newblock {\em J. Phys. G}, 41:074001, 2014.

\bibitem{carl2016}
B.~D.~Carlsson $et$ $al.$.
\newblock {Uncertainty analysis and order-by-order optimization of chiral
  nuclear interactions}.
\newblock {\em Phys. Rev. X}, 6(1):011019, 2016.

\bibitem{PDG2020}
P.A.Zyla $et$ $al.$.
\newblock {Review of Particle Physics}.
\newblock {\em Prog.Theor.Exp.Phys.}, 2020(8):083C01, 2020.

\bibitem{meng2019}
Lu~Meng and Shi-Lin Zhu.
\newblock Light pseudoscalar meson and doubly charmed baryon scattering lengths
  with heavy diquark-antiquark symmetry.
\newblock {\em Phys. Rev. D}, 100:014006, 2019.

\bibitem{eber2002}
D.~Ebert, R.~N. Faustov, V.~O. Galkin, and A.~P. Martynenko.
\newblock Mass spectra of doubly heavy baryons in the relativistic quark model.
\newblock {\em Phys. Rev. D}, 66:014008, 2002.

\bibitem{guo2017}
Zhi-Hui Guo.
\newblock Prediction of exotic doubly charmed baryons within chiral effective
  field theory.
\newblock {\em Phys. Rev. D}, 96:074004, 2017.

\end{thebibliography}

\end{document}